\documentclass[useAMS,usenatbib]{mn2e}
\pdfoutput=1
\usepackage{graphicx}
\usepackage{dcolumn}
\usepackage{bm}
\usepackage{amssymb,amsmath,bm}  
\usepackage{color}
\usepackage[colorlinks,linkcolor=red,citecolor=blue,urlcolor=blue ]{hyperref}
\usepackage{multirow}
\usepackage[utf8]{inputenc}
\usepackage{balance}
\usepackage{enumitem}
\usepackage{lipsum}
\usepackage{ulem}
\usepackage{xcolor}

\newcommand{\nv}{\hat{\boldsymbol{\theta}}}
\newcommand{\wisc}{WI$\times$SC}

\def\bH{b_{\scriptscriptstyle\rm H}}
\def\planck{{\it Planck\/}}
\def\citejap#1{\citeauthor{#1}\ \citeyear{#1}}

\title[Tomographic measurement of the intergalactic gas pressure through galaxy-tSZ cross-correlations]{Tomographic measurement of the intergalactic gas pressure through galaxy-tSZ cross-correlations}
\author[Koukoufilippas et al.]{Nick Koukoufilippas$^1$, David Alonso$^1$, Maciej Bilicki$^2$, John A. Peacock$^3$\\
                      $^{1}$Department of Physics, University of Oxford, Keble Road, Oxford, OX1 3RH, United Kingdom\\
                      $^{2}$Center for Theoretical Physics, Polish Academy of Sciences, al. Lotnik\'ow 32/46, 02-668, Warsaw, Poland\\
                      $^{3}$Institute for Astronomy, University of Edinburgh, Royal Observatory, Edinburgh, EH9 3HJ, United Kingdom
                      }

\begin{document}
  \date{\today}
  \pagerange{1--19} \pubyear{2020}
  \maketitle

\begin{abstract}
    We cross-correlate maps of the thermal Sunyaev-Zeldovich (tSZ) Compton-$y$ parameter published by \planck\ with the projected distribution of galaxies in a set of low-redshift tomographic bins. We use the nearly full-sky 2MASS Photometric Redshift and WISE $\times$ SuperCOSMOS public catalogues, covering the redshift range $z\lesssim0.4$. Our measurements allow us to place constraints on the redshift dependence of the mass-observable relation for tSZ cluster count analyses in terms of the so-called `hydrostatic mass bias' parameter $1-\bH$. These results can also be interpreted as measurements of the bias-weighted average gas pressure $\langle bP_e\rangle$ as a function of redshift, a quantity that can be related to the thermodynamics of gas inside haloes and used to constrain energy injection processes. We measure $1-\bH$ with $\sim6\%$ precision in 6 equispaced redshift bins, and find no evidence for a redshift-dependent mass bias parameter, in agreement with previous analyses. Our mean value of $1-\bH = 0.75\pm0.03$ is also in good agreement with the one estimated by the joint analysis of \planck\ cluster counts and CMB anisotropies calibrated with CMB lensing. Our measurements of $\langle bP_e\rangle$, at the level of $\sim10\%$ in each bin, are the most stringent constraints on the redshift dependence of this parameter to date, and agree well both with previous measurements and with theoretical expectations from shock-heating models.
\end{abstract}

\begin{keywords}
  cosmology: large-scale structure of the Universe, observations -- methods: data analysis
\end{keywords}

\section{Introduction}\label{sec:intro}
  Modern observational cosmology has reached a stage where constraints from most current and future datasets are limited by astrophysical systematic uncertainties, i.e. our lack of detailed understanding of the small-scale physics behind the luminous components of the Universe, galaxies and gas \citep[e.g.][]{2011MNRAS.415.3649V,2011MNRAS.417.2020S,2014JCAP...04..028F,2015MNRAS.454.2451E,2015MNRAS.454.1958M,2015JCAP...12..049S,2019MNRAS.488.1652H,2019JCAP...03..020S,2019OJAp....2E...4C}. The limiting factor for cosmological constraints from cluster number counts is the uncertainty in the mass-observable relation \citep{2009ApJ...692.1060V,2010ApJ...722.1180V,2011ApJ...732...44S,2013JCAP...07..008H,2014MNRAS.440.2077M,2014A&A...571A..20P,2015MNRAS.446.2205M,2016A&A...594A..24P,2016ApJ...832...95D,2019ApJ...878...55B}. Cosmic shear measurements are strongly affected by sub-grid baryonic physics, and by uncertainties in the processes by which galaxies acquire correlated intrinsic alignments \citep[e.g.][]{2001MNRAS.320L...7C,2002MNRAS.332..788M,2004PhRvD..70f3526H,2017MNRAS.465.2033J,2018PhRvD..98d3528T,2018arXiv181206076H,2018ARA&A..56..393M,2019MNRAS.tmp.2187S}. Understanding the impact of baryons on the matter power spectrum requires better knowledge of the distribution of gas in haloes. Finally, even though the clustering pattern of galaxies is one of the cosmological observables with the highest signal-to-noise ratio, the cosmological constraints that can be extracted from it are severely limited by our incomplete understanding of the galaxy-dark matter connection \citep[see e.g.][and references therein]{2018ARA&A..56..435W}. Even the analysis of the temperature anisotropies in the Cosmic Microwave Background (CMB), arguably the cleanest cosmological observable, are currently limited by the impact of astrophysical foregrounds on small scales \citep{2014ApJ...782...74H,2017JCAP...06..031L,2019arXiv190712875P}.
  
  The CMB secondary anisotropies, in particular the thermal and kinetic Sunyaev-Zeldovich effects \citep[tSZ and kSZ respectively,][]{1972CoASP...4..173S}, as well as the gravitational lensing of CMB photons, have gained popularity as a means to address these issues \citep{2017JCAP...11..040B,2019BAAS...51c.297B}. These effects are relatively clean probes of some of the physical quantities that need to be understood in order to mitigate the impact of astrophysical uncertainties: the matter and gas densities, the gas pressure, and the velocity field. Since these observables are also sensitive to cosmology, their combination with large-scale structure data can be extremely powerful at disentangling astrophysical and cosmological parameters. This has been explored by a large number of groups: CMB lensing data have been used to constrain the bias of different tracers of the matter distribution \citep[e.g.][]{2019MNRAS.485.1720H,2018JCAP...04..053A,2019PhRvD.100b3541A,2018MNRAS.481.1133P}, as well as to calibrate the measurement of galaxy shapes in cosmic shear analyses \citep{2019PhRvD.100b3541A}. The tSZ effect has been used in cross-correlation with galaxy clustering and weak lensing data to determine the physical properties of the diffuse gas, as well as to potentially improve constraints on the amplitude of matter fluctuations \citep{2014PhRvD..89b3508V,2014JCAP...02..030H,2015JCAP...09..046M,2017MNRAS.471.1565H,2017ApJ...845...71A,2018PhRvD..97f3514A,2018MNRAS.480.3928M,2019A&A...624A..48D,2019MNRAS.483..223T,2019arXiv190306654T,2019arXiv190413347P,2019arXiv190707870M}. The kSZ has been used in cross-correlation with galaxy clustering to constrain the growth of structure and the gas density profile around haloes \citep{2016PhRvD..93h2002S,2016A&A...586A.140P,2016PhRvL.117e1301H,2016MNRAS.461.3172S,2017JCAP...03..008D}.
  
  In this work we focus on the cross-correlation between galaxy clustering data and maps of the tSZ Compton-$y$ parameter, making use of existing data from the \planck\ collaboration \citep{2016A&A...594A..22P} and a set of photometric galaxy surveys \citep{2014ApJS..210....9B,2016ApJS..225....5B}. The availability of redshift information allows us to place constraints on the cosmic evolution of the thermal gas pressure and the thermal energy in haloes, as well as to examine any redshift dependence of the mass bias for tSZ cluster studies. This is a relevant topic given the current mild tensions between SZ cluster number counts and CMB primary anisotropies \citep{2016A&A...594A..24P,2016ApJ...832...95D,2019ApJ...878...55B,2019MNRAS.489..401Z}, which could be caused by the assumptions made to model the $y$-mass relation. Conversely, under the assumption that the relation between mass and gas pressure is well understood, the tSZ effect can be thought of as a mass tracer, which can be used to break degeneracies with the galaxy bias parameter -- we however leave this analysis for future work.

  This paper is structured as follows: Section \ref{sec:theory} presents the theoretical background used here to describe our two main observables, the projected overdensity of galaxies and the Compton-$y$ parameter, as well as their cross-correlation. Section \ref{sec:data} presents the datasets used for our analysis. The methods used to analyse these data are described in Section \ref{sec:methods}, and Section \ref{sec:results} presents the results. We summarise our conclusions in Section \ref{sec:conclusion}.

\section{Theory}\label{sec:theory}
  Our work focuses on the cross-correlation of the projected galaxy overdensity in consecutive redshift bins, $\delta_g$, and maps of the tSZ Compton-$y$ parameter.
    
  Here, $\delta_g$ is simply the overdensity in the number of galaxies integrated over a redshift bin:
  \begin{equation}
    \delta_g(\nv)=\int dz\,\phi_g(z)\,\Delta_g(\chi(z)\,\nv),
  \end{equation}
  where $\nv$ is a unit vector on the sphere, $\chi(z)$ is the comoving radial distance to redshift $z$ and $\phi_g(z)$ is the normalised galaxy redshift distribution in the bin. $\Delta_g({\bf x})=n_g({\bf x})/\bar{n}_g-1$ is the 3D galaxy overdensity, where $n_g$ is the galaxy number density.
    
  The Compton-$y$ parameter in turn is given by \citep{1972CoASP...4..173S}:
  \begin{equation}\label{eq:compty}
    y(\nv)=\frac{\sigma_{\scriptscriptstyle \rm T}}{m_ec^2}\int \frac{d\chi}{(1+z)} P_e(\chi\nv),
  \end{equation}
  where $P_e=n_e\,T_e$ is the electron pressure ($n_e$ and $T_e$ are the electron density and temperature respectively), $\sigma_{\scriptscriptstyle\rm T}$ is the Thomson scattering cross-section, and $m_e$ is the electron mass. For a fully ionised gas, the electron pressure is directly related to the total thermal gas pressure through $P_{\rm th}=P_e\,(8-5Y)/(4-2Y)$, where $Y$ is the helium mass fraction (with $Y\simeq0.24$).
    
  \subsection{Projected fields and angular power spectra}\label{ssec:theory.cls}    
    Both $\delta_g$ and $y$ can be described as a projected quantity $u(\nv)$ related to a three-dimensional field $U({\bf r})$ through some radial kernel $W_u(\chi)$:
    \begin{equation}
      u(\nv)=\int d\chi\,W_u(\chi)\,U(\chi\nv).
    \end{equation}
    Any projected quantity can be decomposed into its spherical harmonic coefficients $u_{\ell m}$, the covariance of which is the so-called angular power spectrum $\langle u_{\ell m}v^*_{\ell' m'}\rangle\equiv C^{uv}_\ell\delta_{\ell\ell'}\delta_{mm'}$.

    The angular power spectrum can be related to the 3D power spectrum of the associated 3D fields $P_{UV}$ via\footnote{Note that Eq.\!~\ref{eq:cllimber} is only valid in the {\sl Limber approximation} \citep{1953ApJ...117..134L,1992ApJ...388..272K}, which is accurate for broad radial kernels. This approximation is adequate for the redshift distributions and scales used here.}:
    \begin{equation}\label{eq:cllimber}
      C_\ell^{uv} = \int d\chi \frac{W_u(\chi)W_v(\chi)}{\chi^2}\,P_{UV}\left(k=\frac{\ell+1/2}{\chi},z(\chi) \right).
    \end{equation}
    Here, the 3D power spectrum is analogously defined as the variance of the Fourier-space 3D quantities:
    \begin{equation}
      \left\langle U({\bf k})V^*({\bf k}')\right\rangle = (2\pi)^3\,\delta({\bf k}-{\bf k}')\,P_{UV}(k).
    \end{equation}
    We model the 3D power spectrum using the halo model, which we describe in the next section.

    In this formalism, the 3D quantities associated with $\delta_g(\nv)$ and $y(\nv)$ are the 3D overdensity $\Delta_g({\bf x})$ and the electron pressure $P_e({\bf x})$ respectively. The associated radial kernels are:
    \begin{equation}
      W_g(\chi)=\frac{H(z)}{c}\,\phi_g(z),\hspace{12pt}W_y(\chi)=\frac{\sigma_{\scriptscriptstyle\rm T}}{m_ec^2}\frac{1}{1+z},
    \end{equation}
    where $H(z)$ is the expansion rate.

  \subsection{Halo model predictions}\label{ssec:theory.hm}
    The halo model describes the spatial fluctuations of any quantity in terms of the contributions of all dark matter haloes, under the assumption that all matter in the Universe is contained in those haloes. We only quote here the final results regarding the halo model prediction for power spectra, and refer the reader to \cite{2000MNRAS.318..203S,2000MNRAS.318.1144P,2002PhR...372....1C} for further details.
    
    Let $U(r|M)$ be the profile of a given quantity as a function of the comoving distance $r$ to the centre of a halo of mass $M$, and let $U(k|M)$ be its Fourier transform:
    \begin{equation}
      U(k|M)\equiv4\pi \int_0^\infty dr\,r^2\,\frac{\sin(kr)}{kr}U(r|M).
    \end{equation}
    The halo model prediction for the cross-power spectrum $P_{UV}$ then consists of two contributions, the so-called 1-halo term and 2-halo term:
    \begin{equation}
      P_{UV}(k)=P^{1h}_{UV}(k)+P^{2h}_{UV}(k).
    \end{equation}
    Each of these can be estimated in terms of the Fourier-space profiles as:
    \begin{align}
      &P^{1h}_{UV}(k)=\int dM\,\frac{dn}{dM}\,\langle U(k|M)\,V(k|M)\rangle,\\
      &P^{2h}_{UV}(k)=\langle bU\rangle\,\langle bV\rangle\,P_L(k),\\
      &\langle bU\rangle(k)\equiv\int dM\frac{dn}{dM}\,b_h(M)\,\langle U(k|M)\rangle. \label{eq:hm_bias}
    \end{align}
    Here, $P_L(k)$ is the linear matter power spectrum, $dn/dM$ is the halo mass function (comoving density of haloes per unit halo mass) and $b_h(M)$ is the halo bias.

    It is important to note that the halo model is inaccurate in the range of scales corresponding to the transition between the 1-halo and 2-halo-dominated regimes. This is a well-known effect \citep{2015MNRAS.454.1958M}, and we correct for it here simply by multiplying all halo-model power spectra by a universal scale-dependent factor, given by the ratio between the revised {\sl Halofit} prediction for the matter power spectrum of \cite{2012ApJ...761..152T} and the pure halo-model prediction for the same quantity
    \begin{equation}
      R(k)\equiv\frac{P_{\rm Halofit}(k)}{P_{\rm halo\,model}(k)}.
    \end{equation}
    
    \subsubsection{Galaxies and the halo occupation distribution}\label{sssec:theory.hm.hod}
      To model the galaxy overdensity, $\Delta_g$, we use a Halo Occupation Distribution (HOD) model \citep{2002ApJ...575..587B,2005ApJ...633..791Z,2013MNRAS.430..725V}, as prescribed by \cite{2011ApJ...736...59Z}. The HOD models the galaxy content of dark matter haloes as being made up of central and satellite galaxies. Centrals lie at the centre of the halo, while satellites are distributed according to a profile $u_s(r|M)$. Haloes can have zero or one central, and the mean number of centrals for a halo of mass $M$ is modelled as a smoothed step function
      \begin{equation}
        \langle N_c(M)\rangle=\frac{1}{2}\left[1+{\rm erf}\left(\frac{\log(M/M_{\rm min})}{\sigma_{\rm lnM}}\right)\right].
      \end{equation}
      In our fiducial scenario we assume that satellites can only be formed if a halo has a central and has a mass larger than some threshold $M_0$. In that case, the average number of satellites follows a power law of the form:
      \begin{equation}\label{eq:hod1}
        \langle N_s(M)\rangle=N_c(M)\,\Theta(M-M_0)\,\left(\frac{M-M_0}{M_1'}\right)^{\alpha_s}.
      \end{equation}
      For simplicity, we fix $M_0$ to $M_{\rm min}$, $\sigma_{\rm lnM}=0.15$ and $\alpha_s$=1, as in \cite{2018MNRAS.473.4318A}, leaving only two free parameters: $M_{\rm min}$ and $M_1'$. Coupling $M_0$ and $M_{\rm min}$ allows for all haloes containing a central to also contain satellites, and conversely, for all haloes containing satellites to necessarily contain a central. This assumption breaks down in cases such as recent major mergers, where centrals may not immediately be established.
      
      Besides their mean values, we also need to specify the statistics of $N_c$ and $N_s$. Following standard practice \citep{2013MNRAS.430..725V}, we assume $N_c$ to have a Bernoulli distribution with probability $p=\langle N_c\rangle$, and $N_s$ to be Poisson-distributed.

      Putting everything together, the moments of the galaxy overdensity Fourier profile are \citep[e.g. see section 2.2 of][]{2013MNRAS.430..725V}:
      \begin{align}
        &\langle u_g(k)\rangle=\bar{n}_g^{-1}\left[\langle N_c\rangle	+\langle N_s\rangle\,u_s(k)\right],\\
        &\langle |u_g(k)|^2\rangle=\bar{n}_g^{-2}\left[\langle N_s\rangle^2u_s^2(k)+2\langle N_s\rangle u_s(k)\right],
      \end{align}
      where the mean number density $\bar{n}_g$ is
      \begin{equation}
        \bar{n}_g\equiv\int dM\,\frac{dn}{dM}\left(\langle N_c\rangle+\langle N_s\rangle\right),
      \end{equation} 
      where we have suppressed the mass dependence of all quantities for brevity.

      Finally, we assume that the satellites follow the matter distribution, and therefore $u_s(k|M)$ is given by a truncated Navarro, Frenk \& White profile \citep{1996ApJ...462..563N}:
      \begin{align}
        u_s(k|M)=&\left[\log(1+c_\Delta)-\frac{c_\Delta}{(1+c_\Delta)}\right]^{-1}\\\nonumber
               &\left[\cos(q)\left({\rm Ci}((1+c_\Delta)q)-{\rm Ci}(q)\right)\right.\\\nonumber
               &\left.+\sin(q)\left({\rm Si}((1+c_\Delta)q)-{\rm Si}(q)\right)\right.\\\nonumber
               &\left.-\sin(c_\Delta q)/(1+c_\Delta q)\right],
      \end{align}
      where $q\equiv kr_\Delta/c_\Delta$, $r_\Delta$ and $c_\Delta$ are the halo radius and concentration defined in Section \ref{sssec:theory.hm.cm}, and $\{{\rm Ci}, {\rm Si}\}$ are the cosine and sine integrals.
      
    \subsubsection{tSZ and pressure profiles}\label{sssec:theory.hm.pe}
      In order to describe the electron pressure in a halo, we use the generalised NFW profile (GNFW) described in \cite{2010A&A...517A..92A} and used in the \planck\ tSZ cluster analysis \citep{2016A&A...594A..24P}. This profile takes the form:
      \begin{equation}
        P_e(r)=P_*\,p(r/r_{500c}),
      \end{equation}
      where $r_{500c}$ is the cluster radius enclosing an overdensity of 500 times the critical density (see Section \ref{sssec:theory.hm.cm}). The normalisation $P_*$ is given by
      \begin{equation}\label{eq:arnaud_norm}
        P_*=6.41\,\left(1.65\,\, {\rm eV}\,{\rm cm}^{-3}\right)h_{70}^{8/3}
        \left(\frac{h_{70}(1-\bH)M_{500c}}{3\times10^{14}{\rm M_\odot}}\right)^{0.79},
      \end{equation}
      where $h_{70}=H_0/(70\ {\rm km}\,{\rm s}^{-1}\,{\rm Mpc}^{-1})$, and $M_{500c}$ is the halo mass enclosed by $r_{500c}$. The GNFW form factor is
      \begin{equation}
        p(x)=(c_P x)^{-\gamma}\left[1+(c_P x)^\alpha\right]^{(\gamma-\beta)/\alpha},
      \end{equation}
      with $(\alpha,\beta,\gamma,c_P)=(1.33,4.13,0.31,1.81)$. We must note that other pressure profiles have been proposed in the literature (e.g. \citejap{2012ApJ...758...75B}), but we choose this parameterisation in order to be able to relate our measurement of $(1-\bH)$ to the results of \cite{2016A&A...594A..24P}.
      
      The quantity $1-\bH$ in Eq.\!~\ref{eq:arnaud_norm} parameterises our lack of knowledge about the precise relation between mass and pressure in clusters. This factor is also commonly referred to as the `hydrostatic bias', since it was originally defined to account for the fraction of halo mass not in hydrostatic equilibrium missed by X-ray observations. Since this parameter also encapsulates other sources of bias in the X-ray-based mass estimates, we instead refer to it as the {\sl mass bias}. Numerical simulations have constrained the mass deficit to be around 20\% (i.e. $1-\bH\simeq0.8$: \citejap{2012ApJ...758...74B}; \citejap{2014ApJ...782..107N}), although it is known that a smaller value is necessary in order to fully reconcile CMB primary constraints and SZ cluster counts \citep{2016A&A...594A..24P}. This is a central point of our discussion in Sections \ref{sec:results} and \ref{sec:conclusion}.

      Within the halo model description, and assuming a log-normal $y$-mass relation, the pressure profile cumulants are given by:
      \begin{align}
        &\langle u_y(k|M)\rangle=P_e(k),\\
        &\langle u_y^2(k|M)\rangle=P_e^2(k)\,e^{\sigma_{\ln Y}^2},
      \end{align}
      where $P_e(k)$ is the Fourier transform of the GNFW profile and $\sigma_{\ln Y}=0.173\pm0.023$ is the intrinsic logarithmic scatter in the $y$-mass relation \cite{2016A&A...594A..24P}. Note that, since we do not make use of the tSZ auto-correlation, we do not use the second order cumulant of the pressure profile in our analysis. However, since we analyse the galaxy-tSZ correlation, we need to model the covariance between the galaxy overdensity and pressure profiles. For simplicity, we adopt a one-parameter model:
      \begin{equation}
        \langle u_y(k|M) u_g(k|M)\rangle = (1+\rho_{yg})\langle u_g(k|M)\rangle \langle u_y(k|M)\rangle,
      \end{equation}
      where the free parameter $\rho_{yg}$ determines the sign of the correlation between galaxy abundance and pressure. Marginalising over this parameter has the added advantage of removing any sensitivity of our final constraints to $1-\bH$ on the details of the cross-spectrum model in the 1-halo regime, where both parameters are completely degenerate.

      It is also worth exploring the halo model bias (Eq.\!~\ref{eq:hm_bias}) for the Compton-$y$ parameter. At $k\rightarrow0$ it is given by:
      \begin{align}\nonumber
        \langle bP_e\rangle&=\int dM\,\frac{dn}{dM}\,b_h(M)\,\int_0^\infty dr\,4\pi r^2\,P_e(r|M)\\\label{eq:by}
               &=\int dM\,\frac{dn}{dM}\,b_h(M)\,E_T(M),
      \end{align}
      where $E_T(M)$ is the thermal energy in a halo of mass $M$ \citep{2017MNRAS.467.2315V,2019arXiv190413347P}. A measurement of $\langle bP_e\rangle$ can therefore be related to the thermodynamics of gas inside haloes. We also express our measurements in terms of this parameter in Section \ref{sec:results}.
      
    \subsubsection{Concentration-mass relation and mass definitions}\label{sssec:theory.hm.cm}
      Halo radii $r_\Delta$ are usually defined as the size of the sphere containing a given mass $M_\Delta$:
      \begin{equation}
        M_\Delta = \frac{4\pi}{3}\rho_*(z) \; \Delta \; r^3_\Delta.
      \end{equation}
      Common choices for $\rho_*$ are the critical density, $\rho_c=3H^2(z)/8\pi G$, or the matter density, $\rho_M(z)$. The spherical overdensity parameter, $\Delta$, is usually chosen within the range $\sim(200,500)$ and sometimes defined as the quantity yielding the virial radius in the spherical top-hat collapse model, $\Delta_v$ \citep{1998ApJ...495...80B}.

      Ideally we would like to use the same mass definition (i.e. choice of $\Delta$ and $\rho_*$) for the mass function, the mass-concentration relation $c_\Delta(M)$ and the calibrated pressure profile. Unfortunately while the GNFW profile is calibrated to $\Delta_{500c}$ (where $c$ denotes critical density), the mass functions of \cite{2008ApJ...688..709T,2010ApJ...724..878T} are only provided for $\rho_M$-based mass definitions, and the concentration-mass relation of \cite{2008MNRAS.390L..64D} was only estimated for $\Delta=200$ (for critical and matter densities) and for $\Delta=\Delta_v$. To overcome this issue, we follow the procedure used by \cite{2016A&A...594A..24P} and \cite{2018MNRAS.477.4957B}: our baseline mass definition is $\rho_c$-based with $\Delta=500$, as used by \cite{2010A&A...517A..92A}. At each redshift, we translate this into a $\Delta$ value for a $\rho_M$-based definition, which we use to compute the mass function from the parameterisations of \cite{2008ApJ...688..709T,2010ApJ...724..878T}. We also re-derive the concentration-mass relation of \cite{2008MNRAS.390L..64D} for a $\rho_c$-based $\Delta=500$ from their $\Delta=200$ parameterisation by integrating the NFW profile to the corresponding halo radius. Within the redshifts covered by our analysis we find that this is well fit by:
      \begin{equation}
        c_{500c}(M,z)= A\,(M/M_{\rm pivot})^B\,(1+z)^C,
      \end{equation}
      with $M_{\rm pivot}=2.7\times10^{12}\,{\rm M_\odot}$ and $(A,B,C)=(3.67,\,-0.0903,\,-0.51)$.

\section{Data}\label{sec:data}
  \subsection{The Compton-$y$ map}\label{ssec:data.y}
    \begin{figure}
      \centering
      \includegraphics[width=0.5\textwidth]{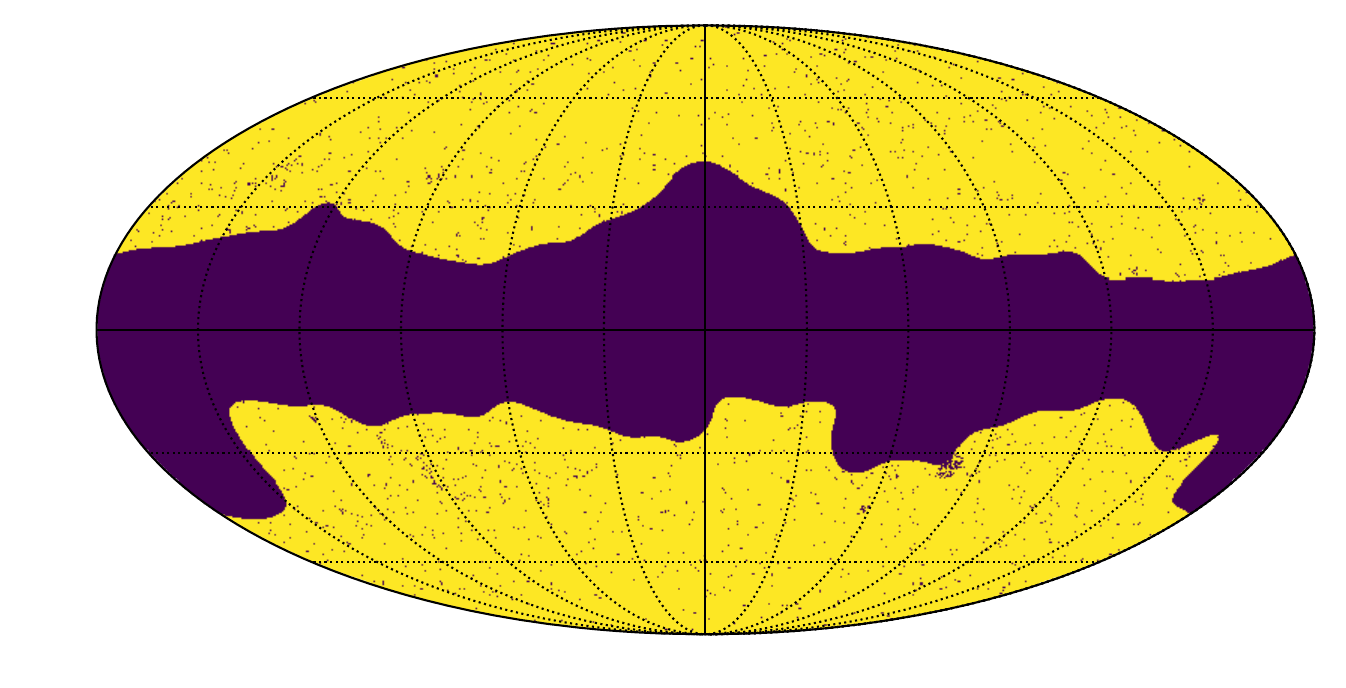}
      \includegraphics[width=0.5\textwidth]{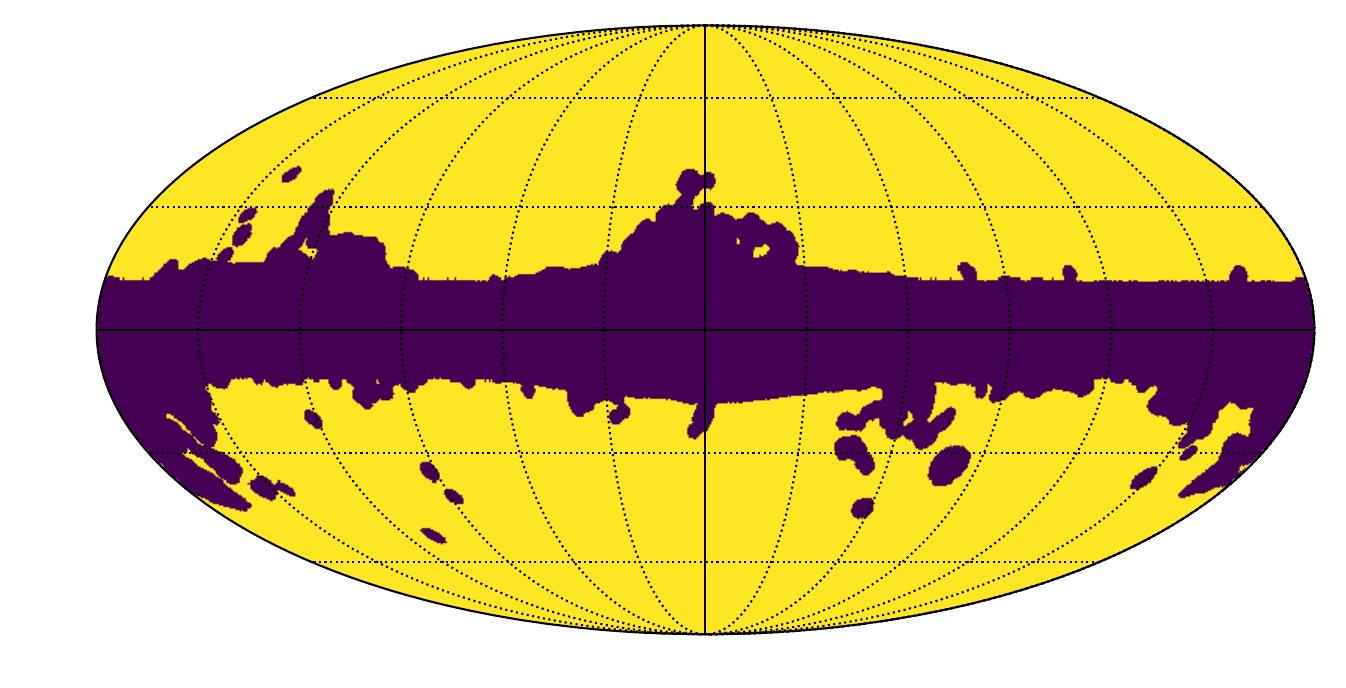}
      \caption{{\sl Top:} sky mask used for the Compton-$y$ map, corresponding to a sky fraction $f_{\rm sky}=0.59$. {\sl Bottom:} mask used for the 2MPZ and \wisc~ galaxy surveys, corresponding to a sky fraction $f_{\rm sky}=0.68$. The product of both masks leaves a usable sky fraction $f_{\rm sky}\simeq0.58$.
      }
      \label{fig:msk}
    \end{figure}
    We make use of the Compton-$y$ parameter maps made public by the \planck\ collaboration \citep{2016A&A...594A..22P}. These maps were generated using different flavours of the Internal Linear Combination method \citep{2004ApJ...612..633E,2008arXiv0811.4277V}. In a simplified description, the ILC technique selects the linear combination of all frequency channels that preserves the spectrum of the source one wishes to map, minimising the map-level variance. The refined versions of the ILC method used by \planck\ further optimise the linear weights on different scales and different regions of the map, and project out sources with known spectra that are likely to cause significant contamination. In particular, \planck\ has released two $y$ maps, extracted using the MILCA \citep{2013A&A...558A.118H} and NILC \citep{2011MNRAS.410.2481R} variations of the ILC technique. Both methods deproject CMB contamination through its well-known spectrum, but differ on the methods used to calculate the optimal scale-dependent and spatially-varying linear weights.
    
    The MILCA and NILC maps have been found to be in good agreement in different studies, although the NILC map has a higher noise level on large scales \citep{2016A&A...594A..22P}. We thus use the MILCA map as our fiducial Compton-$y$ map, but repeat our analysis on the NILC map as part of our systematics analysis. We use a fiducial mask for the $y$ maps based on a combination of the \planck\ 60\% Galactic mask and the union of the HFI and LFI point source masks (see top panel of Fig.\!~\ref{fig:msk}).
    
    Finally, in order to evaluate the level of contamination from extragalactic dust in the $\delta_g$-$y$ correlation, we make use of the HFI 545 GHz map, as described in Section \ref{sssec:results.syst.y}.
    
  \subsection{2MPZ and \wisc}\label{ssec:data.g1}
    \begin{figure}
      \centering
      \includegraphics[width=0.5\textwidth]{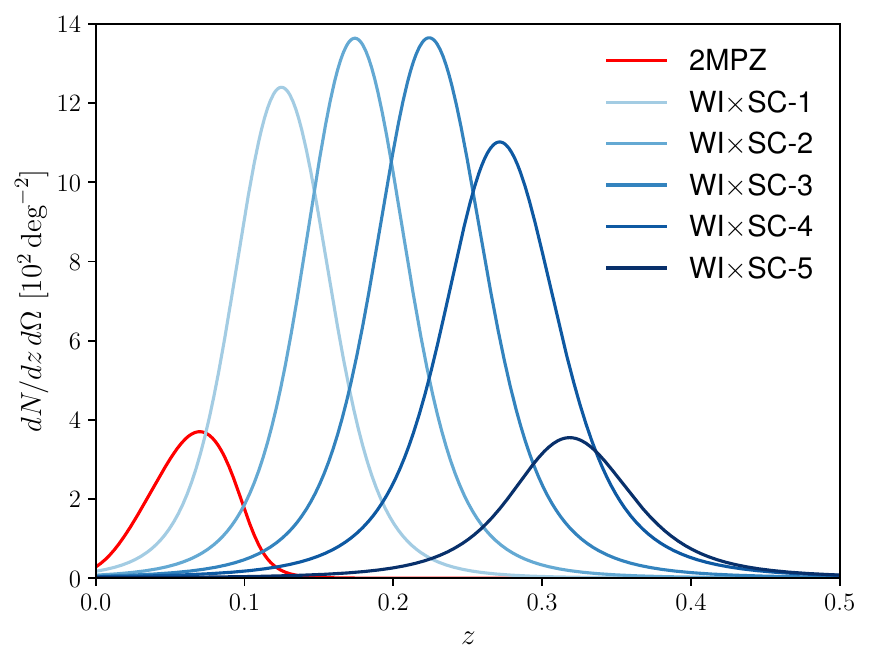}
      \caption{Fiducial redshift distributions of the different galaxy samples used in this analysis. See Table \ref{tab:z_bins} for further details.}
      \label{fig:dndz}
    \end{figure}
    \begin{table}
      \begin{center}
        \begin{tabular}{l|cccc}
          \hline
          Sample & $[z_{{\rm ph},i},z_{{\rm ph},f}]$ & $\bar{z}$ & $\bar{n}_g\,[{\rm deg}^{-2^{\phantom{2}}}]$ & $\ell_{\rm max}$\\[1ex]
          \hline
          2MPZ    & N.A.         & 0.07 &  25.5 &  280 \\
          \wisc-1 & $[0.1,0.15]$ & 0.13 & 106   &  540 \\
          \wisc-2 & $[0.15,0.2]$ & 0.18 & 126   &  745 \\
          \wisc-3 & $[0.2,0.25]$ & 0.23 & 136   &  945 \\
          \wisc-4 & $[0.25,0.3]$ & 0.27 & 118   & 1130 \\
          \wisc-5 & $[0.3,0.35]$ & 0.32 & 41    & 1310 \\
          \hline
        \end{tabular}
        \caption{Galaxy samples used in this analysis, corresponding to the full 2MPZ survey and five tomographic redshift bins of the \wisc~ survey. The second, third and fourth columns list the photometric redshift interval defining the sample, its mean redshift and its number density respectively. The largest multipole used in the analysis of each sample (corresponding to a comoving scale of $k_{\rm max}\sim 1\,{\rm Mpc}^{-1}$) is shown in the last column.}\label{tab:z_bins}
      \end{center}
    \end{table} 
    We make use of two low-redshift photometric redshift (photo-$z$) catalogues, the 2MASS Photometric Redshift catalogue \citep[2MPZ,][]{2014ApJS..210....9B} and the WISE $\times$ SuperCOSMOS catalogue \citep[\wisc,][]{2016ApJS..225....5B}. Both samples were created by cross-matching full-sky imaging surveys, and photo-$z$'s were subsequently computed for all the included sources. Their broad photometric coverage, together with adequate spectroscopic calibration data allows for well-constrained photo-$z$'s, with minimal mean bias, relatively low scatter and a small number of outliers.
    
    2MPZ was constructed by cross-matching the extended-source catalogue from the 2 Micron All-Sky Survey \citep[2MASS,][]{2006AJ....131.1163S,2000AJ....119.2498J} with the photographic plates of SuperCOSMOS \citep{2001MNRAS.326.1295H,2016MNRAS.462.2085P} and the photometry of the Wide-field Infrared Survey Explorer (WISE: \citejap{2010AJ....140.1868W}). After applying an apparent magnitude cut $K_s<13.9$ (using Vega system) to achieve uniformity, 2MPZ includes over 940,000 sources observed in 8 bands: SuperCOSMOS's optical $(B,R,I)$, 2MASS's near-infrared $(J,H,K_s)$ and WISE's mid-infrared $(W1,W2)$. Photometric redshifts were extracted using the neural network code {\tt ANNz} \citep{2004PASP..116..345C} trained on a large spectroscopic sample from overlapping surveys. The resulting photo-$z$'s have a typical error $\sigma_z\simeq0.015$, and the sample has a median redshift $z\simeq0.08$ (see also \citealt{2018MNRAS.476.1050B} for a more detailed analysis of the 2MPZ photo-$z$ properties). As demonstrated in Section \ref{sec:results}, due to the low redshift of this sample, the cross-correlation with the $y$ map is dominated by the 1-halo term for 2MPZ, and little is gained by sub-dividing it into narrower redshift bins. Therefore we use 2MPZ as a single tomographic sample.
    
    A deeper sample is obtained by ignoring the 2MASS data and cross-matching WISE and SuperCOSMOS only. After removing the sources already contained in 2MPZ, the resulting catalogue, \wisc, is $\sim3$ times deeper, contains $\sim20$ million sources and reaches up to redshift $z\sim0.4$, with a median redshift of $\sim0.2$. The photometric redshifts are less accurate, given the poorer photometric coverage $(B,\,R,\,W1,\,W2)$, with a mean error $\sigma_z/(1+z)\simeq0.035$. We divide the \wisc~ sample into 5 redshift bins, corresponding to photo-$z$ intervals of equal width $\delta z_{\rm photo}=0.05$ in the range $0.10<z_{\rm photo}<0.35$. Details regarding each of these redshift bins are given in Table \ref{tab:z_bins}, and the corresponding redshift distributions are shown in Fig.\!~\ref{fig:dndz}. We estimate fiducial redshift distributions for each tomographic sample as described in \cite{2018MNRAS.481.1133P}, and we discuss the marginalisation over uncertainties in these distributions in Section \ref{sssec:methods.syst.photoz}.

    Both 2MPZ and WISC suffer from different levels of contamination from Galactic and observational systematics. The most relevant Galactic systematic is star contamination, particularly for \wisc~ \citep{2019JCAP...08..037X}. Besides avoiding regions of high dust and star contamination using the sky mask described in \cite{2018MNRAS.481.1133P} (see bottom panel of Fig.\!~\ref{fig:msk}), we correct for the effects of stellar contamination by correcting for a smooth non-linear relation between galaxy and star density (also described in \citejap{2018MNRAS.481.1133P}). After doing so, three potential sources of systematic uncertainty remain: residual stellar contamination, modulation of the galaxy density due to Galactic dust reddening and modulation due to zero-point fluctuations in the photographic plates used by SuperCOSMOS. We address the first two (dust and stars) by deprojecting them at the map level as described in Section \ref{sssec:methods.syst.deproj}. We address the contamination from fluctuations in the SuperCOSMOS plates by modelling it at the power spectrum level, which we describe in Section \ref{sssec:methods.syst.plates}.
    
    We depart from the approach used in \cite{2018MNRAS.481.1133P} in that we do not additionally employ the SDSS photometric catalogue \citep{2016MNRAS.460.1371B} as a galaxy tracer. At redshifts $z<0.35$, 2MPZ and \wisc\ have wider-angle coverage than SDSS, which results to higher signal-to-noise ratio of the $y\times \delta_g$ cross-correlation; therefore SDSS could only prove useful at redshifts $z>0.35$. This is, however, where the Dark Energy Survey have already provided constraints \citep{2019arXiv190413347P}, offering in particular well-constrained redshift distributions for their mass tracers (luminous red galaxies). In the case of the SDSS photo-$z$ sample, redshifts are difficult to calibrate in the $z>0.35$ regime. This is because there are no sufficiently complete overlapping wide-angle spectroscopic datasets in order to do it robustly for the purposes of our work (see Section \ref{sssec:methods.syst.photoz}). This is also briefly discussed in Section 2.1 of \cite{2018MNRAS.481.1133P}, although the uncertainties on the SDSS redshift distributions were of minor importance for the final results there. This is not any more the case in this work, as we have verified.

\section{Methods}\label{sec:methods}
  \subsection{Estimating power spectra}\label{ssec:methods.cls}
    We measure all auto- and cross-power spectra between the different redshift bins and the $y$ maps using the pseudo-$C_\ell$ estimator \citep[e.g.][]{2002ApJ...567....2H} as implemented in the {\tt NaMaster} code\footnote{\url{https://github.com/LSSTDESC/NaMaster}.} \citep{2019MNRAS.484.4127A}. Details about the method can be found in these references, but we provide a brief description here for completeness. For an incomplete sky coverage, a given field observed on the sphere, $\tilde{u}(\nv)$, can be modelled as a product of the true underlying field $u$, and a sky mask $w$
    \begin{equation}
      \tilde{u}^{\rm obs}(\nv) = w(\nv)\,u(\nv).
    \end{equation}
    In the simplest scenario, the mask $w$ is simply a binary map ($w=0$ or 1) selecting the pixels in the sky that have been observed. More generally, $w$ can be designed to optimally up- or downweight different regions in an inverse-variance manner. Through the convolution theorem, the spherical harmonic transform of the observed field is a convolution of the harmonic transforms of the true field and the mask. Provided the mask and true field are uncorrelated, this then translates into a similar result for the ensemble average of the observed power spectra $\tilde{C}^{uv}_\ell$:
    \begin{equation}
      \tilde{C}^{uv}_\ell = \sum_{\ell'}\,M^{uv}_{\ell \ell'}\, C^{uv}_{\ell'},
    \end{equation}
    where $C^{uv}_\ell$ is the true underlying power spectrum. $M^{uv}_{\ell \ell'}$ is the so-called mode-coupling matrix, which depends solely on the masks of both fields, and which can be computed analytically. Roughly speaking, the pseudo-$C_\ell$ approach is then based on estimating $M^{uv}$ and inverting it to yield an unbiased estimate of the power spectrum.
    
    For the galaxy auto-correlation, the pseudo-$C_\ell$ method requires an additional step of subtracting the shot noise bias. We do so analytically following the approach described in Section 2.4.2 of \cite{2019MNRAS.484.4127A} with a local noise variance given by $\sigma_n^2=1/\bar{n}_\Omega$, where $\bar{n}_\Omega$ is the mean surface number density in units of inverse steradians.
    
    All maps were generated and operated on using the {\tt HEALPix} pixelisation scheme \citep{2005ApJ...622..759G} with resolution parameter $N_{\rm side}=512$, corresponding to a pixel size $\theta_{\rm pix}\sim7'$ ($\ell \sim 1535$).

  \subsection{Covariance matrices}\label{ssec:methods.cov}
    We combine two different methods to estimate the power spectrum covariance matrix. 
    
    We make a first estimate of the covariance using the jackknife resampling method. We divide the common footprint covered by the masks of both the galaxy overdensity and Compton-$y$ maps into $N_{\rm JK}$ regions of roughly equal area. We mask each region in turn and compute the power spectrum  using the remaining available footprint. The covariance matrix is then estimated as:
    \begin{equation}\label{eq:cov_jk}
      {\rm Cov}\left(C^{uv}_\ell,C^{wz}_{\ell'}\right)=\frac{N_{\rm JK}-1}{N_{\rm JK}}\sum_{n=1}^{N_{\rm JK}}\Delta C^{uv,(n)}_\ell\,\Delta C^{wz,(n)}_{\ell'},
    \end{equation}
    where $\Delta C^{uv,(n)}_\ell$ is the difference between the power spectrum estimated when removing the $n$-th jackknife region and the power spectrum averaged over all jackknife regions. We use $N_{\rm JK}=461$ jackknife regions defined as {\tt HEALPix} pixels with resolution $N_{\rm side}=8$.

    Although the jackknife method is able to provide an estimate of the size of the power spectrum uncertainties in a model independent way, it has some drawbacks. There are only so many jackknife regions of reasonable size that can be selected; therefore the estimated covariance is noisy at some level. Typically, the number of independent realisations used to estimate a sample covariance matrix should be at least one order of magnitude larger than the size of the data vector (up to 50 elements in our case). Furthermore, since the footprint associated with the removal of each jackknife region is different from the overall footprint, the method is also not able to recover the mode-coupling associated with the map geometry by construction. In order to verify and improve our estimate of the covariance matrix, we make use of a second analytical estimator.

    We compute the analytical covariance matrix following the methods outlined in \cite{2017MNRAS.470.2100K}. The covariance receives two main additive contributions, from the so-called {\sl disconnected} and {\sl connected} trispectra. The disconnected part is essentially the covariance matrix estimated under the assumption that all fields are Gaussian. In the absence of sky masks, it is given by
    \begin{equation}
      {\rm Cov}^{\rm G}\left(C^{uv}_\ell,C^{wz}_{\ell'}\right)=\delta_{\ell\ell'}\frac{C^{uw}_\ell C^{vz}_\ell+C^{uz}_\ell C^{vw}_\ell}{2\ell+1}.
    \end{equation}
    A sky mask introduces non-zero coupling between different $\ell$ modes. To account for these, we use the  method introduced in \cite{2004MNRAS.349..603E} and implemented in {\tt NaMaster}, which has been shown to be an excellent approximation for large-scale structure data \citep{2019arXiv190611765G}.

    We compute the connected (i.e. non-Gaussian) contribution to the covariance matrix using the halo model as the 1-halo trispectrum \citep{2002MNRAS.336.1256K}, given by:
    \begin{align}\nonumber
      {\rm Cov}^{\rm NG}\left(C^{uv}_\ell,C^{wz}_{\ell'}\right)=\int &d\chi\,\frac{W_u(\chi)W_v(\chi)W_w(\chi)W_z(\chi)}{4\pi f_{\rm sky}\,\chi^6}\times\\\label{eq:cov_ng}
      &T^{1h}_{uvwz}\left(k=\frac{\ell+1/2}{\chi}\right),
    \end{align}
    where
    \begin{equation}\nonumber
      T^{1h}_{uvwz}(k, k^\prime)\equiv\int dM\frac{dn}{dM}\langle U(k|M) V(k|M) W(k^\prime|M) Z(k^\prime|M)\rangle.
    \end{equation}
    Here we have used the notation introduced in Section \ref{ssec:theory.cls} for the radial kernels ($W_u(\chi)$) and Fourier-space halo profiles ($U(k|M)$) of the different projected fields. The total covariance matrix is simply given by ${\rm Cov}={\rm Cov}^{\rm G}+{\rm Cov}^{\rm NG}$. Note that estimating the connected term requires the use of the best-fit halo model parameters which we do not know a priori. In order to circumvent this issue we proceed as in \cite{2018MNRAS.473.4318A} and estimate the covariance matrix in a two-step process, where we first obtain best-fit parameters by minimising a $\chi^2$ that uses only the Gaussian covariance with power spectra computed directly from the data, and we then use those parameters to calculate the non-Gaussian contribution (as well as to recalculate the Gaussian part using the best-fit prediction for the power spectra).
    \begin{figure}
      \centering
      \includegraphics[width=0.5\textwidth]{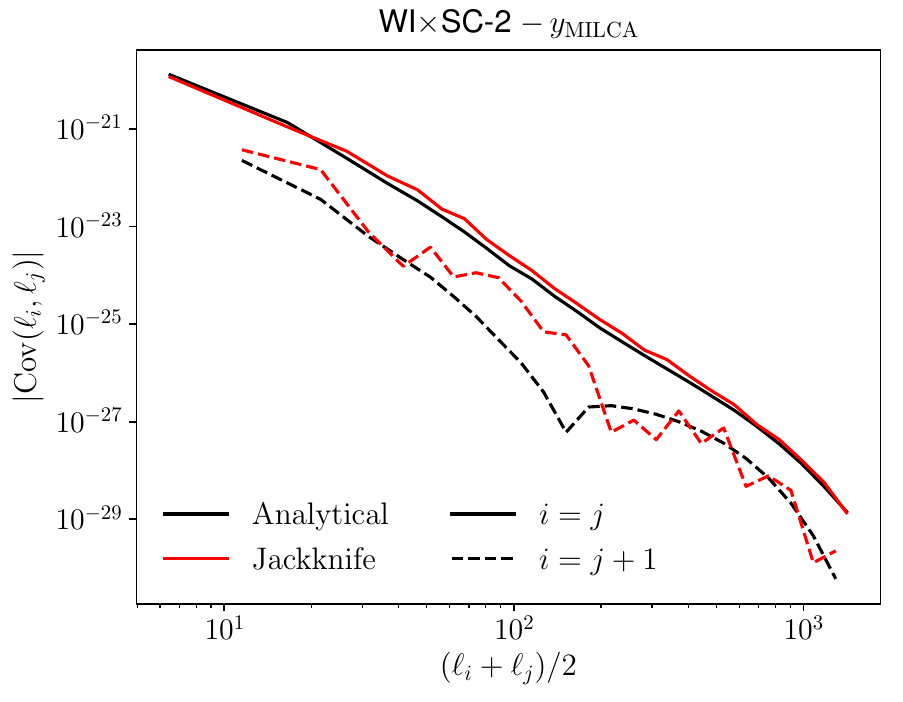}
      \caption{Diagonals of the covariance matrix for the $y\times \delta_g$ cross-correlation in the second \wisc~redshift bin. Black and red lines show the analytical and jackknife covariances respectively. The solid and dashed lines show the zeroth-order ($i=j$) and first-order ($i=j+1$) diagonals respectively. We find compatible results from both methods.}
      \label{fig:covdiag}
    \end{figure}

    Fig.\!~\ref{fig:covdiag} shows the diagonal of the covariance matrix for one of the galaxy-tSZ power spectra estimated using these two methods. We find that both estimators are in good agreement with each other. We construct our final fiducial covariance matrix using both estimators: in order to ensure that we recover realistic error bar sizes, we use the variance estimated from the jackknives, and then combine it with the correlation matrix estimated analytically. This ensures that our estimator accounts for the coupling between different modes caused by survey geometry and non-Gaussianities while avoiding the statistical noise in the jackknife estimator. The final covariance is therefore:
    \begin{equation}
      {\rm Cov}_{ij} = {\rm Cov}^{\rm ana}_{ij} \sqrt{\frac{{\rm Cov}^{\rm JK}_{ii}\,{\rm Cov}^{\rm JK}_{jj}}{{\rm Cov}^{\rm ana}_{ii}\,{\rm Cov}^{\rm ana}_{jj}}},
    \end{equation}
    where ${\rm Cov}^{\rm JK}$ and ${\rm Cov}^{\rm ana}$ are the jackknife and analytical covariance matrices respectively. We have verified that the resulting covariance matrices are well behaved (i.e. they are invertible, and their eigenvalues have a reasonable dynamical range).

  \subsection{Systematics treatment}\label{ssec:methods.syst}
    \subsubsection{Map-level deprojection}\label{sssec:methods.syst.deproj}
      For small levels of contamination, the impact of a given systematic on an observed sky map can be modelled at the linear level:
      \begin{equation}\label{eq:deproj}
        {\bf m}_{\rm obs}={\bf m}_{\rm true} + \epsilon\,{\bf t}.
      \end{equation}
      Here ${\bf m}_{\rm obs}$ and ${\bf m}_{\rm true}$ are vectors corresponding to the observed sky map and the true underlying quantity we wish to map respectively, ${\bf t}$ is a template map describing the contaminant (e.g. a map of the Galactic dust fluctuations), and $\epsilon$ is an unknown amplitude. In order to fully account for the effects of this contaminant one has to build a likelihood for ${\bf m}_{\rm obs}$ using the Eq. \ref{eq:deproj} as a model (together with a model for ${\rm m}_{\rm true}$) and marginalise over $\epsilon$. As shown in \cite{2017MNRAS.465.1847E} and \cite{2019MNRAS.484.4127A}, within the pseudo-$C_\ell$ framework, this can be done exactly by projecting ${\bf m}_{\rm obs}$ onto the subspace perpendicular to ${\bf t}$ or, in other words, by `deprojecting' ${\bf t}$. The loss of modes due to deprojection then needs to be taken into account when estimating the power spectrum, which can be done analytically.

\begin{figure}
        \centering
        \includegraphics[width=0.48\textwidth]{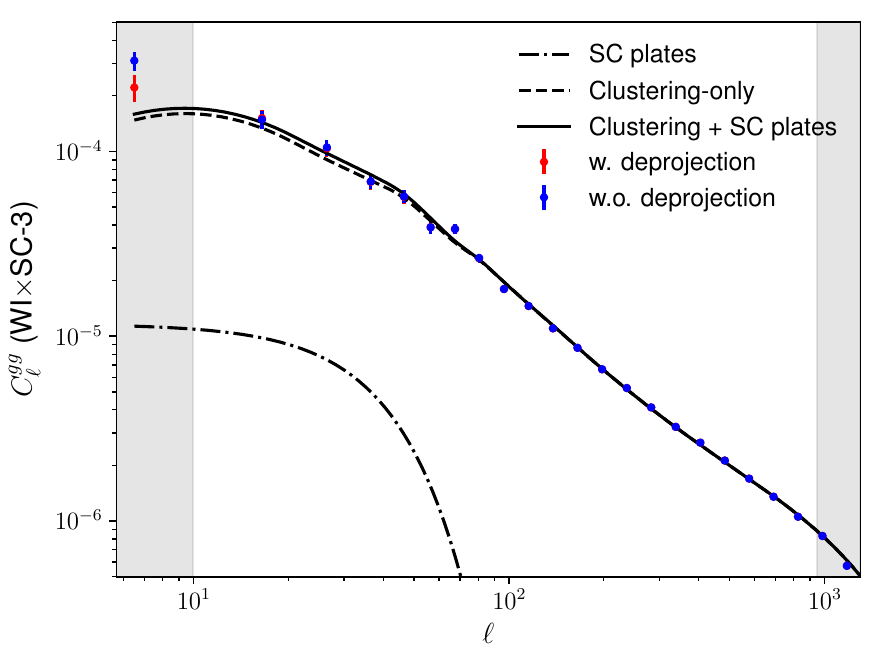}
        \caption{Summary of the different sources of systematic contamination affecting the galaxy auto-correlations for the particular case of the third \wisc~bin. The red and blue points show the measurements before and after deprojecting a dust and a star template. The dot-dashed line shows the best-fit contamination from SuperCOSMOS plate fluctuations that explains the residuals of the data with respect to the best-fit HOD-only model (dashed line). The total combined prediction is shown as the solid line. The light grey bands show our scale cuts.}
        \label{fig:clsyst}
      \end{figure}

      For our analysis, we deproject two systematic templates from the galaxy and $y$ maps. We create a reddening template using the dust map of \cite{1998ApJ...500..525S}, and remove its associated contamination from the $y$ maps and from all the galaxy overdensity maps. We also generate a star density template from the WISE data, and remove its associated fluctuations from all the galaxy overdensity maps. The effects of this deprojection are illustrated in Fig.\!~\ref{fig:clsyst}.

    \subsubsection{$C_\ell$-level deprojection}\label{sssec:methods.syst.plates}
      We model the systematic fluctuations in the number density of sources in the \wisc~ sample caused by variations in the zero-point of the SuperCOSMOS photographic plate exposures as Gaussian random variations within the footprint of each exposure with a variance $\sigma^2_{\rm plate}$:
      \begin{equation}
        \delta_{\rm plate}(\nv) = \sum_p \delta_p\,S_{\rm plate}\,W(\nv-\nv_p),
      \end{equation}
      where the sum runs over all exposures, $\delta_p$ is the fluctuation in exposure $p$ (with $\langle\delta_p^2\rangle=\sigma^2_{\rm plate}$), $S_{\rm plate}$ is the footprint area of each exposure and $W(\nv)$ is the plate window function. Assuming a roughly homogeneous coverage of the sky, the power spectrum of these fluctuations is given by
      \begin{equation}
        C_\ell^{\rm plate}=S_{\rm plate}\sigma_{\rm plate}^2\left|W_\ell\right|^2,
      \end{equation}
      where $W_\ell$ is the harmonic transform of the plate window function. SuperCOSMOS used photographic plates covering an area of $5^\circ\times5^\circ$. The corresponding window function can be roughly approximated as $|W_\ell|^2=\exp[-(\ell\,\theta_{\rm plate})^2/12]$, where $\theta_{\rm plate}=5\pi/180$ for 5-degree plates.
      
      The contamination from plate fluctuations can therefore be accounted for as an additive contribution to the model describing the galaxy auto-spectrum, proportional to a template $T_\ell\equiv|W_\ell|^2$ with a free amplitude $A\equiv S_{\rm plate}\sigma_{\rm plate}^2$ that is marginalised over. Since $A$ is a linear parameter, this can be done analytically in a pre-processing step by modifying the inverse covariance matrix of the galaxy auto-spectra as follows \citep{1992ApJ...398..169R}:
      \begin{equation}
        {\sf Cov}^{-1}\hspace{6pt}\rightarrow\hspace{6pt} {\sf Cov}^{-1}-\frac{{\sf Cov}^{-1}{\bf T}\cdot{\bf T}^T{\sf Cov}^{-1}}{{\bf T}^T{\sf Cov}^{-1}{\bf T}},
      \end{equation}
      where ${\bf T}$ is a vector containing the template $T_\ell$ for all the scales used in the analysis. The dot-dashed line in Fig.\!~\ref{fig:clsyst} shows the shape of the plate template $T_\ell$ normalised by the best-fit value of the parameter $A$ that explains the residuals of the data with respect to the best-fit HOD-only model (shown as a dashed line). The plate fluctuations contribute to $\sim10\%$ of the signal on scales $\ell\lesssim 30$.

    \subsubsection{Scale cuts}\label{sssec:methods.syst.scales}
      After deprojecting dust and stars at the map level, and the imprint of the SuperCOSMOS plates at the power spectrum level, we still observe an unacceptably large amount of power on scales $\ell<10$ in all galaxy auto-correlations involving \wisc. These may be due to residual star contamination that is not simply removed with a linear template \citep{2019JCAP...08..037X}, or inaccuracies in our treatment of the SuperCOSMOS fluctuations. To avoid biasing our results we therefore remove the lowest bandpower from all galaxy auto-correlations involving \wisc.
      
      On small scales, the simple halo model prescription used to describe the non-linear power spectra and covariance matrices may not be sufficiently accurate. Therefore we impose a cut on angular multipoles $\ell$ larger than the typical physical scale of a halo, $\ell_{\rm max}=k_{\rm max}\bar{\chi}-1/2$, where $k_{\rm max}=1\,{\rm Mpc}^{-1}$, and $\bar{\chi}$ is the mean comoving radial distance in each redshift bin. The corresponding values of $\ell_{\rm max}$ in each bin are given in Table \ref{tab:z_bins}.
      
    \subsubsection{Redshift distribution uncertainties}\label{sssec:methods.syst.photoz}
      For each redshift bin we estimate fiducial redshift distributions using the method described in \cite{2018MNRAS.481.1133P}. In short, a true-redshift distribution is estimated from the distribution of photometric redshifts for a given bin using a model for the conditional photo-$z$ distribution:
      \begin{equation}\label{eq:nz_calc}
        p(z)=\int dz_{\rm photo}\,p(z|z_{\rm photo})\,p(z_{\rm photo}),
      \end{equation}
      where the model for $p(z|z_{\rm photo})$ was calibrated using overlapping spectroscopic data. Although the spectroscopic coverage of 2MPZ and \wisc~ is high compared to other photometric redshift surveys, the resulting redshift distributions are not infinitely precise, and therefore we need to account for any uncertainty in them that could affect our measurement.

      In particular, our constraints on $1-\bH$ are very sensitive to the width of the redshift distributions. This is because wider redshift distributions, which lead to lower projected clustering amplitudes, affect the galaxy auto-correlation, $\delta_g\times \delta_g$, while the $y\times\delta_g$ cross-correlation is almost insensitive to the width. We introduce an additional parameter, $w_z$, that stretches the support of the redshift distribution while preserving unit total probability. Concretely, given a fiducial distribution $p_{\rm fid}(z)$ with mean redshift $\bar{z}$, $w_z$ is implemented as
      \begin{equation}
        p(z)\propto p_{\rm fid}\left(\bar{z}+\frac{z-\bar{z}}{w_z}\right),
      \end{equation}
      where the proportionality constant is fixed by making sure that $p(z)$ integrates to unity. We make $w_z$ a free parameter that we marginalise over with a top-hat prior $0.8<w_z<1.2$. This prior is significantly larger than the actual expected uncertainty in the width of the redshift distributions. To be precise, we estimate that the parameters (e.g. Gaussian mean and variance) of the conditional photo-$z$ distribution $p(z|z_{\rm photo})$ used to calculate the true redshift distributions (Eq.\!~\ref{eq:nz_calc}), are known to the level of $1\%$. When propagated into an uncertainty on the width $w_z$ of $p(z)$, this corresponds to an uncertainty of the same order ($\sim0.9\%$). Thus, the margins of the assumed top-hat priors are wide enough to encompass any lack of precision in the number density distribution.
      
      Another possible source of systematic uncertainty is the effect of biased photometric redshifts (i.e. where $\langle z|z_{\rm photo}\rangle\neq z_{\rm photo}$). The {\tt ANNz} method used in \cite{2018MNRAS.481.1133P} should guarantee unbiased photo-$z$'s, but this can never be achieved exactly in practice. It would be possible to account for this form of uncertainty by treating $\bar{z}$ as a free parameter, as is usually done in photometric cosmic shear analyses. Our results, however, are more sensitive to the distribution widths, and we find that this simple parameterisation is able to describe our data sufficiently well. We leave a more detailed analysis of the associated photometric redshift systematics for future work.

  \subsection{Likelihood}\label{ssec:methods.like} 
  \begin{figure*}
    \centering
    \includegraphics[width=0.8\textwidth]{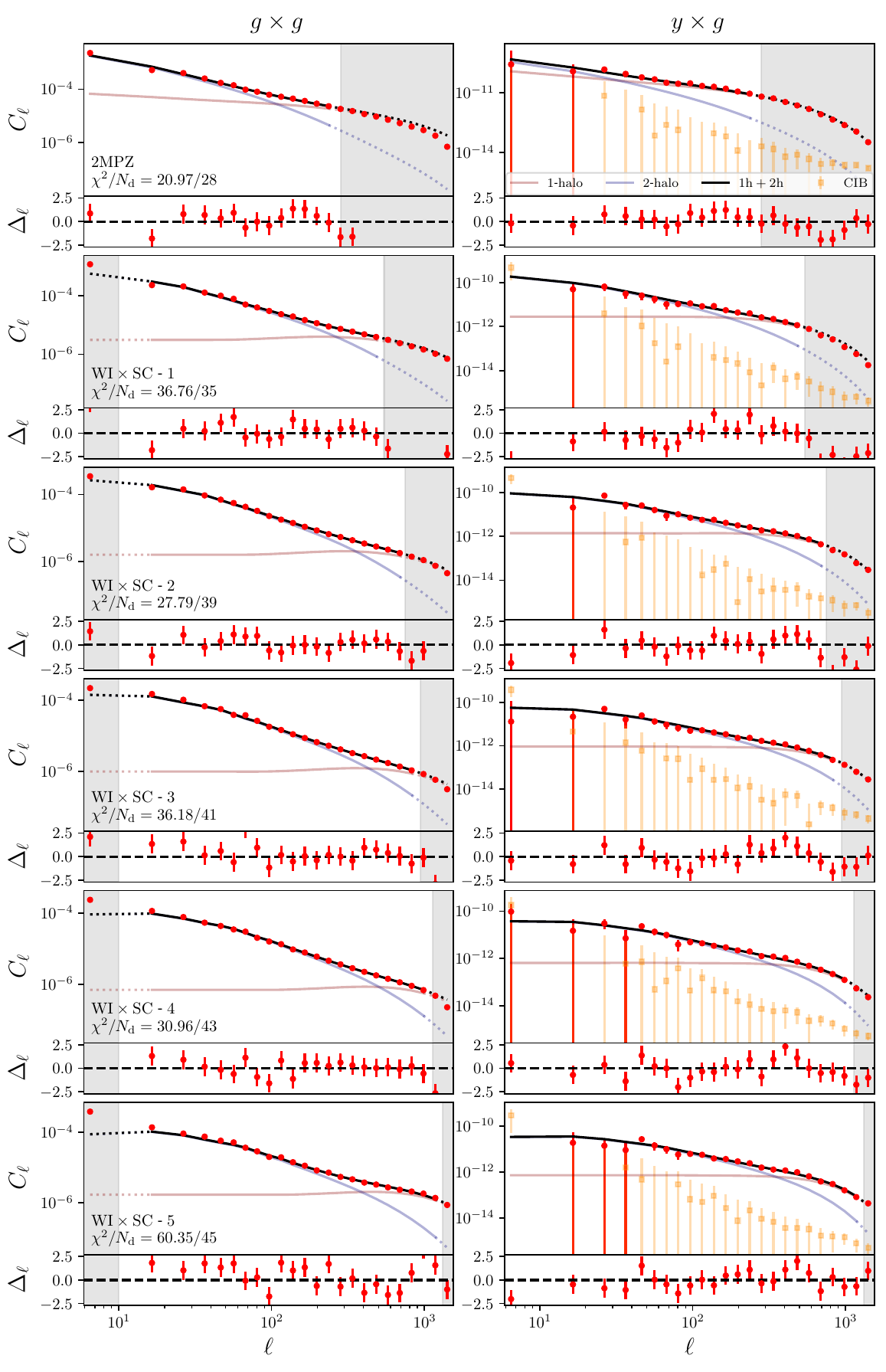}
    \caption{Measured galaxy auto-spectra (left column) and $\delta_g$-$y$ cross-spectra (right column) for the 6 galaxy redshift bins shown in Fig.\!~\ref{fig:dndz} in ascending order. Each panel shows the measurements (red points with error bars) and best-fit predictions (solid black lines) decomposed into their 1-halo and 2-halo contributions (burgundy and blue lines). The right column additionally shows the estimated contamination from extragalactic dust (yellow squares with error bars), which is found to be negligible. The grey bands indicate the scales removed from the analysis. The bottom part of each panel shows the difference between data and theory normalised by the 1$\sigma$ uncertainties.}
    \label{fig:cls}
  \end{figure*}
  \begin{figure*}
    \centering
    \includegraphics[width=0.27\textwidth]{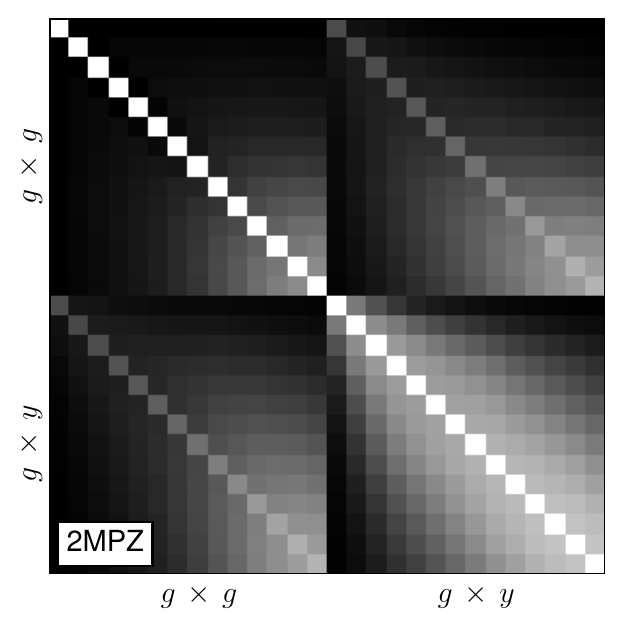}
    \includegraphics[width=0.27\textwidth]{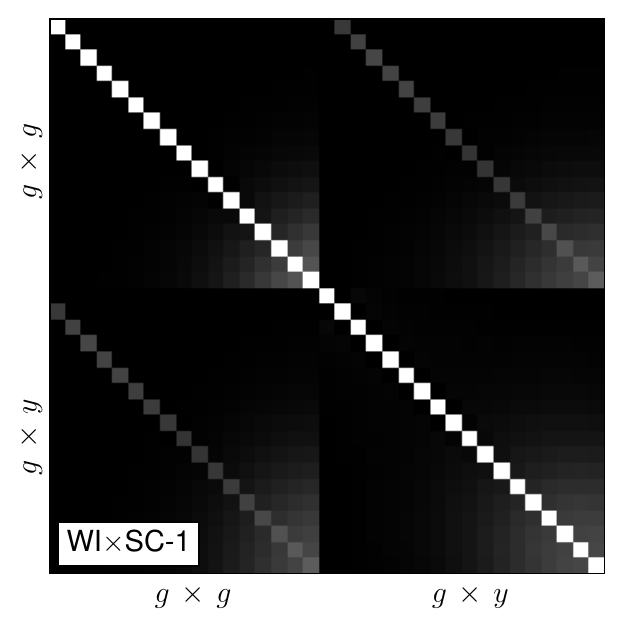}
    \includegraphics[width=0.27\textwidth]{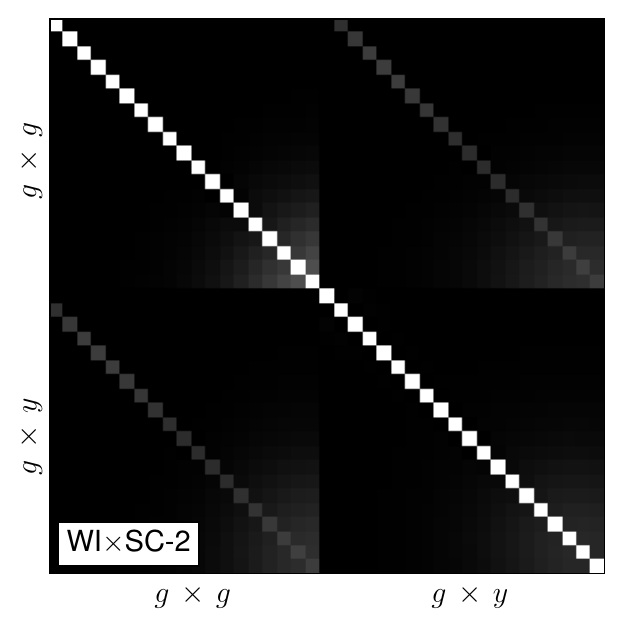}
    \includegraphics[width=0.27\textwidth]{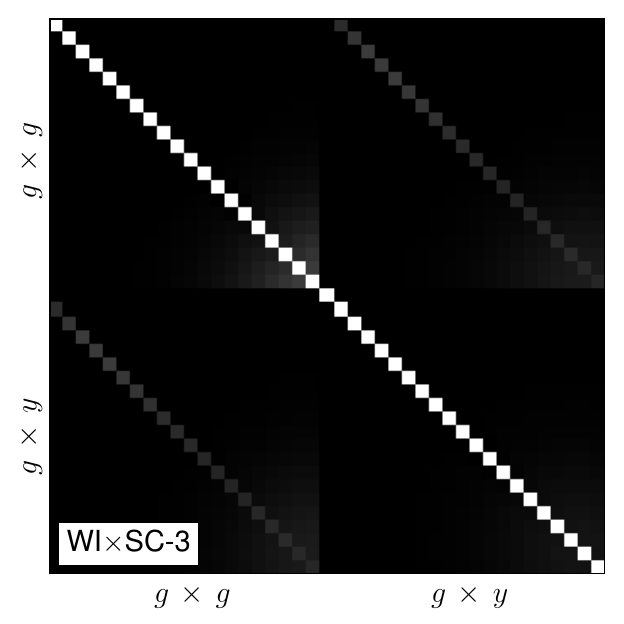}
    \includegraphics[width=0.27\textwidth]{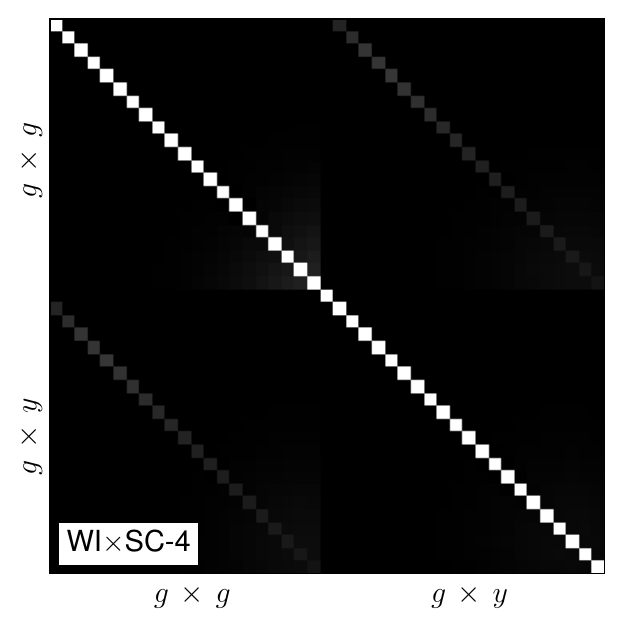}
    \includegraphics[width=0.27\textwidth]{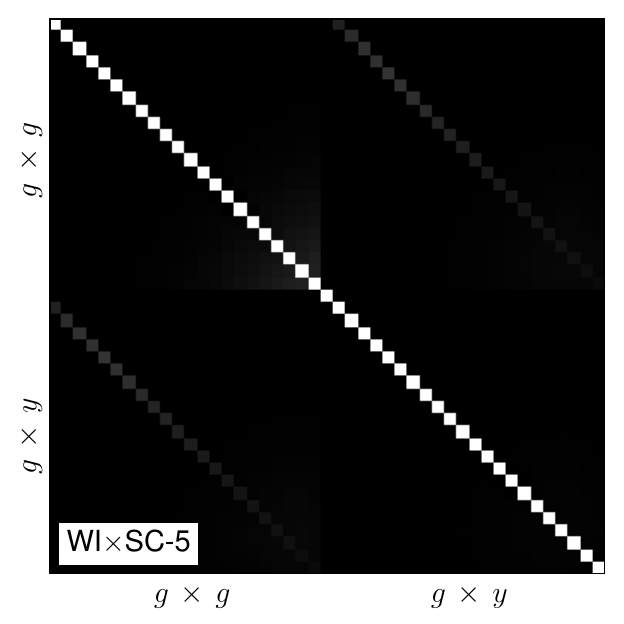}
    \caption{Correlation matrices for the 6 redshift bins used in the analysis. Each covariance matrix consists of 4 sub-matrices, corresponding to the covariances of $C^{gg}_\ell$ and $C^{gy}_\ell$ (block diagonals), as well as their cross-covariance.}
    \label{fig:covs}
  \end{figure*}
  \begin{table}
      \begin{center}
        \begin{tabular}{lll|lll}
          $b$ & $\ell^b_{\rm ini}$ & $\ell^b_{\rm end}$ & $b$ & $\ell^b_{\rm ini}$ & $\ell^b_{\rm end}$ \\[1ex]
          \hline
           1 &   2 &  12 &  13 &  180 &  216\\
           2 &  12 &  22 &  14 &  216 &  258\\
           3 &  22 &  32 &  15 &  258 &  308\\
           4 &  32 &  42 &  16 &  308 &  369\\
           5 &  42 &  52 &  17 &  369 &  441\\
           6 &  52 &  62 &  18 &  441 &  527\\
           7 &  62 &  74 &  19 &  527 &  629\\
           8 &  74 &  88 &  20 &  629 &  752\\
           9 &  88 & 106 &  21 &  752 &  899\\
          10 & 106 & 126 &  22 &  899 & 1074\\
          11 & 126 & 151 &  23 & 1074 & 1284\\
          12 & 151 & 180 &  24 & 1284 & 1535\\
          \hline
        \end{tabular}
        \caption{Power spectrum bandpowers used in this work. The $b$-th bandpower includes all integer multipoles $\ell^b_{\rm ini}<\ell<\ell^b_{\rm end}$. The bandpower edges listed here correspond roughly to linearly-spaced bands between $\ell=2$ and $\ell=52$, and logarithmic bands afterwards.}\label{tab:bpws}
      \end{center}
    \end{table} 
    In order to connect our measurements with the posterior distribution of the model parameters $\vec{\theta}$, we assume that the measured power spectra follow a Gaussian likelihood:
    \begin{equation}
      -2\ln p({\bf d}|\vec{\theta}) = \chi^2\equiv ({\bf d}-{\bf t}(\vec{\theta}))^T {\sf C}^{-1} ({\bf d}-{\bf t}(\vec{\theta})),
    \end{equation}
    where ${\bf d}$ is a vector of power spectrum measurements, ${\bf t}(\vec{\theta})$ is the theory prediction for ${\bf d}$ with parameters $\vec{\theta}$ and ${\sf Cov}$ is the covariance matrix described in the previous sections.

    We explore the likelihood of each redshift bin separately. For each bin, our data vector ${\bf d}$ includes two sets of power spectrum measurements, corresponding to the auto-correlation of the galaxy overdensity and to its cross-correlation with the Compton-$y$ map, ${\bf d}=(C^{gg}_\ell,C^{gy}_\ell)$. We compute all power spectra in the range of multipoles $2<\ell<1535$, binned into the bandpowers described in Table \ref{tab:bpws}. For a given redshift bin, we only include those bandpowers that satisfy the scale cuts described in Section \ref{sssec:methods.syst.scales}.

    For each redshift bin, our theoretical model has five free parameters: $\log_{10}M_{\rm min}/{\rm M_\odot}$, $\log_{10}M_1'/{\rm M_\odot}$, $1-b_{\rm H}$, $\rho_{yg}$, and a nuisance width parameter $w_z$. The first two parameters effectively fit the galaxy bias and small-scale amplitude in the galaxy auto-correlation, while $1-b_{\rm H}$ and $\rho_{yg}$ are then constrained by including the tSZ cross-correlation. We fix all cosmological parameters to the best-fit values in \cite{2018arXiv180706209P}: $(\Omega_c h^2,\Omega_bh^2,h,\sigma_8,n_s)=(0.119,0.0224,0.6766,0.8102,0.9665)$. In our fiducial scenario we adopt the \citet{2008ApJ...688..709T} mass function parameterisation. We have used the halo model bias described in \citet{2010ApJ...724..878T} throughout our analysis; in our fiducial case, and when exploring departures from it. We impose the following top-hat priors on the free parameters:
    \begin{align}
      &10\le\log_{10}M_{\rm min}/{\rm M_\odot}\le16,\\
      &10\le\log_{10}M_1'/{\rm M_\odot}\le16,\\
      &0\le(1-\bH)\le0.99,\\
      &-1\le\rho_{yg}\le1,\\
      &0.8\le w_z\le1.2.
    \end{align}
    We sample the resulting posterior distributions using the Markov chain Monte Carlo method (MCMC) as implemented in the {\tt emcee} software package \citep{2013PASP..125..306F}\footnote{\url{https://emcee.readthedocs.io/en/v2.2.1/}.}. We use the Core Cosmology Library\footnote{\url{https://github.com/LSSTDESC/CCL}} \citep{2019ApJS..242....2C} in our theory calculations.

\section{Results}\label{sec:results}
  \subsection{Power spectra and covariances}\label{ssec:results.cls}
    We estimate the galaxy auto-power spectrum, the galaxy-tSZ cross-spectrum, and their covariance matrix for each of the redshift bins shown in Fig.\!~\ref{fig:dndz} using the methods described in Section \ref{sec:methods}. The resulting measured power spectra and errors are shown in red in Fig.\!~\ref{fig:cls}, together with their best-fit halo model prediction in black, decomposed into its 1-halo and 2-halo contributions (burgundy and blue respectively). The bottom part of each panel shows the residuals with respect to the best-fit prediction normalised by the 1$\sigma$ errors. The grey bands cover the data points not used in the analysis due to scale cuts.
    
    Fig.\!~\ref{fig:covs} shows the correlation matrix of the combined data vector $(C^{gg}_\ell,C^{gy}_\ell)$ for each redshift bin (where the correlation matrix $r_{ij}$ is related to the covariance $C_{ij}$ as $r_{ij}=C_{ij}/\sqrt{C_{ii}C_{jj}}$). At low redshifts, 2MPZ shows strong correlations between different scales, mostly caused by the non-Gaussian contribution to the covariance matrix (Eq.\!~\ref{eq:cov_ng}). These become less relevant at higher redshifts, where non-linear effects are weaker, and where the radial projection pushes the non-linear scale into larger multipole values.
    
    Overall, we find a good agreement between theory and data over the scales used in this analysis. We discuss this agreement and the associated scientific results in the following sections.

  \subsection{Fiducial results}\label{ssec:results.fid}  
    
    \subsubsection{Tomographic measurement of the mass bias}\label{ssec:results.fid.1mb}

      \begin{figure*}
        \centering
        \includegraphics[width=0.7\textwidth]{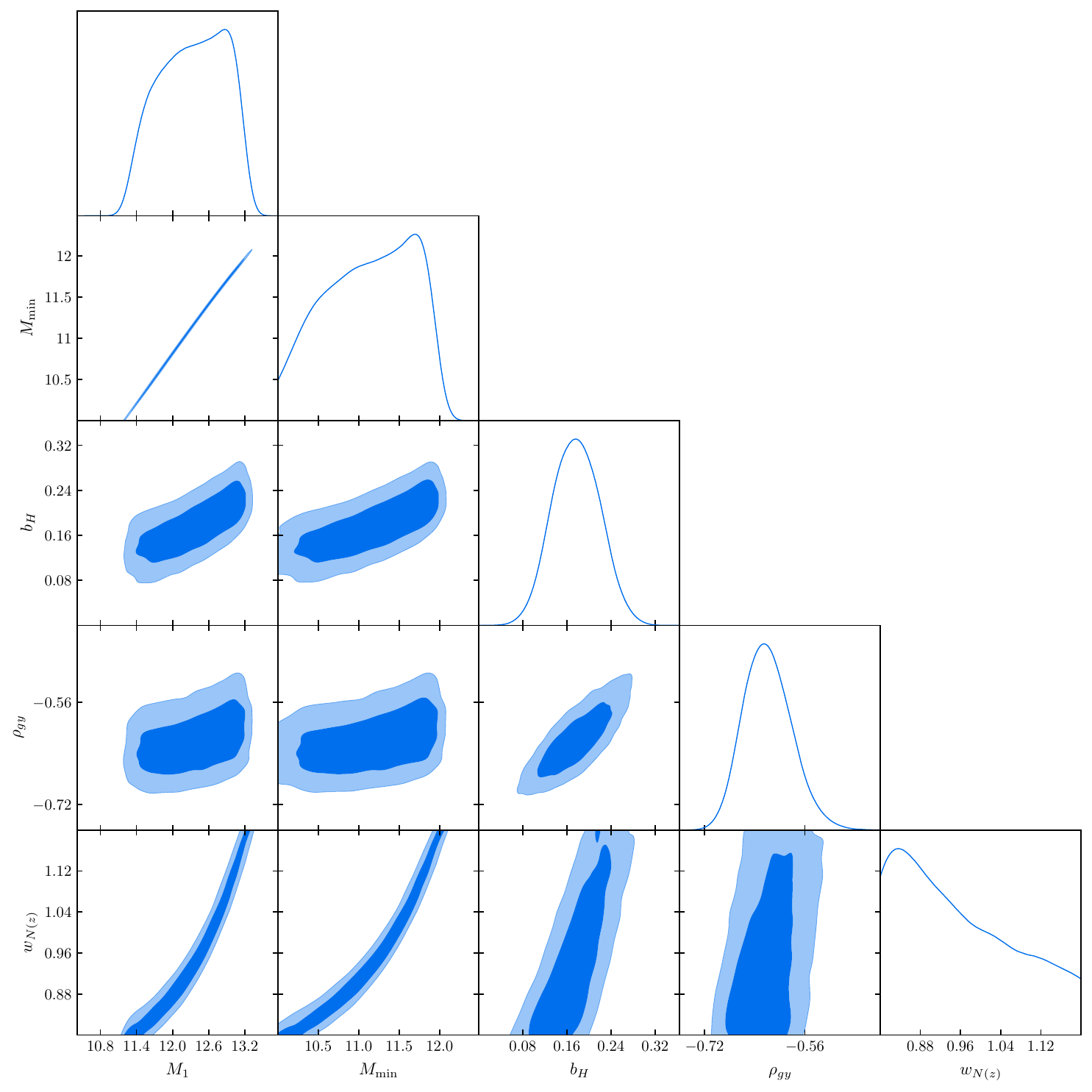}
        \caption{68\% and 95\% contours for the model parameters in the third WI$\times$SC redshift bin.}
        \label{fig:triangle}
      \end{figure*}

      \begin{figure*}
        \centering
        \includegraphics[width=0.7\textwidth]{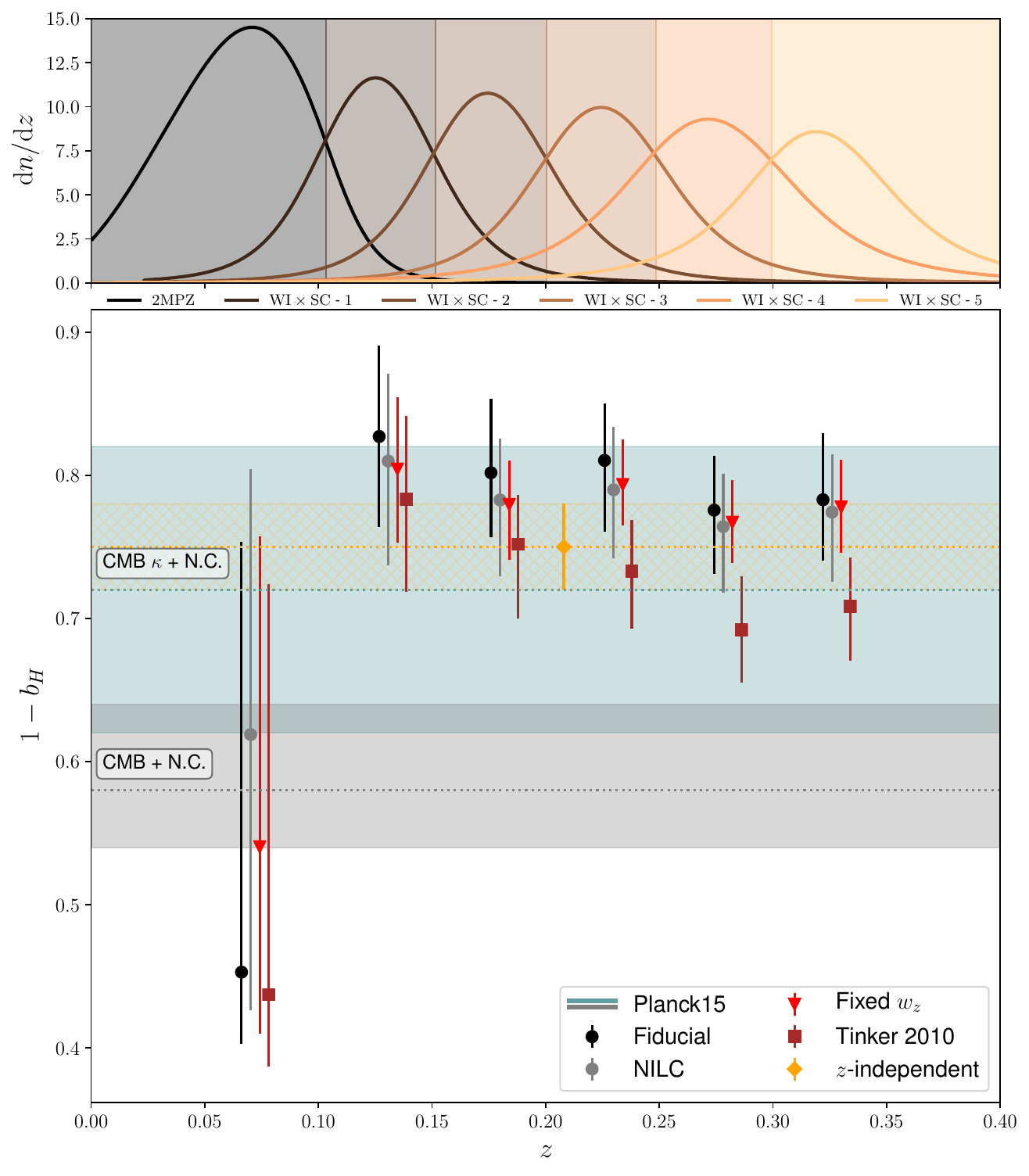}
        \caption{Summary figure showing our constraints on the tSZ mass bias $1-\bH$. Our fiducial constrains are shown as black circles with error bars, and are centred at the mean redshift of each sample. The grey circles show the same constraints found using the NILC $y$ map, which are in good agreement with our fiducial results. The red downward-pointing triangles show the constraints found fixing the photo-$z$ width parameter, $w_z$. They are in good agreement with our fiducial measurements, with $\sim20\%$ smaller error bars. The burgundy squares show the results found using the 2010 Tinker mass function, which are systematically biased low. The orange diamond and the mesh show our combined, redshift-independent constraint on $1-\bH$. For comparison, the grey and turquoise bands show the constraints on the mass bias found by combining cluster counts and CMB primary \citep{2016A&A...594A..24P} as well as the constraints using CMB lensing to calibrate cluster masses \citep{2019MNRAS.489..401Z}, respectively. The top panel shows the normalised redshift distributions for the six redshift bins, for reference.}
        \label{fig:bh}
      \end{figure*}

      \begin{table*}
        \begin{center}
            \begin{tabular}{l|cccccc}
            \hline
            Sample & $\bar{z}$ & $1-\bH\,\,({\rm best\,\,fit})$ & $1-\bH\,\,(68\%\,{\rm C.L.})$ & $\langle bP_e\rangle\,[{\rm meV}\,\,{\rm cm}^{-3^{\phantom{2}}}]$ & $\chi^2/{\rm d.o.f.}$ & ${\rm PTE}(\chi^2)$\\[1ex]
            \hline
            2MPZ    & 0.07     & 0.66  & $0.45^{+0.29}_{-0.10}$     & $0.064^{+0.022^{\phantom{A}}}_{-0.074}$                          & 0.91                  & 0.59               \\
            \wisc-1 & 0.13     & 0.88  & $0.83^{+0.07}_{-0.06}$     & $0.172^{+0.024^{\phantom{A}}}_{-0.028}$                          & 1.23                  & 0.18               \\
            \wisc-2 & 0.18     & 0.84  & $0.80^{+0.05}_{-0.05}$     & $0.187^{+0.025^{\phantom{A}}}_{-0.015}$                          & 0.82                  & 0.76               \\
            \wisc-3 & 0.23     & 0.81  & $0.81^{+0.04}_{-0.04}$     & $0.203^{+0.021^{\phantom{A}}}_{-0.019}$                          & 1.01                  & 0.46               \\
            \wisc-4 & 0.27     & 0.74  & $0.78^{+0.05}_{-0.05}$     & $0.194^{+0.014^{\phantom{A}}}_{-0.026}$                          & 0.82                  & 0.78               \\
            \wisc-5 & 0.32     & 0.75  & $0.78^{+0.04}_{-0.05}$     & $0.225^{+0.025^{\phantom{A}}}_{-0.019_{\phantom{A}}}$            & 1.29                  & 0.09               \\
            \hline
          \end{tabular}
          \caption{Summary table presenting our main results. The first two columns list our 6 tomographic bins and their mean redshift. Columns 3 and 4 show the best-fit value of the mass bias $1-\bH$ and their 1D peak value and 68\% confidence interval respectively. Column 5 shows the peak value and 68\% confidence interval of the bias-averaged thermal pressure (Eq.\!~\ref{eq:by}) in each bin. Finally, columns 6 and 7 show the reduced $\chi^2$ and associated probability to exceed, indicating that we find a good fit in all cases.}\label{tab:results}
        \end{center}
      \end{table*}
      We use the measured power spectra to constrain the free parameters of the model described in Section \ref{sec:theory}, with the main aim of providing an alternative measurement of the mass bias $(1-\bH)$ as a function of redshift. Fig.\!~\ref{fig:triangle} shows an example of the posterior parameter contours for our 5 free parameters, $\{M_{\rm min},M_1,\bH,\rho_{gy},w_z\}$, in the \wisc-3 sample ($z\sim0.23$). As the figure shows, there are strong degeneracies between $M_{\rm min}$, $M_1$ and $w_z$. This is easy to understand, given that all of these parameters affect the overall amplitude of the galaxy auto-correlation. Since we do not probe scales where the 1-halo term is fully resolved, we do not break the degeneracy between $M_{\rm min}$ and $M_1$, which regulate the abundance of centrals and satellites, and therefore the constraint on these parameters mostly comes from the 2-halo amplitude (i.e. the galaxy bias), and the 1-halo shot-noise level. On the other hand, the width of the redshift distribution also has a strong impact on the amplitude of the angular power spectrum at all scales due to projection effects. Since a measurement of the galaxy bias from the galaxy auto-correlation is effectively used to constrain $\bH$ from the galaxy-tSZ cross-correlation, the mass bias also shows some degeneracy with the HOD parameters and $w_z$, albeit at a more moderate level. More interestingly, the mass bias parameter, $\bH$, shows a visible degeneracy with $\rho_{gy}$. This is also expected: the effects of $\rho_{gy}$ and $\bH$ on the 1-halo term of the galaxy-tSZ power spectrum are completely degenerate, and thus a free $\rho_{gy}$ ensures that any information obtained on $\bH$ comes entirely from the 2-halo contribution.
      
      The corresponding constraints on $1-\bH$ for all redshift bins are shown in the main panel of Fig.\!~\ref{fig:bh} as black circles with error bars. These values, which are also given explicitly in column 4 of Table \ref{tab:results}, correspond to the peak of the 1-dimensional distribution of $1-\bH$ and to the equal-probability values encompassing a total probability of 0.68. We report our results on departures from our fiducial model (other data points in Fig.\!~\ref{fig:bh}) in Section \ref{ssec:results.syst}.
      
      In Table \ref{tab:results}, column 3 shows the maximum-likelihood value of $1-\bH$. Columns 6 and 7 of the same table list the reduced $\chi^2$ values for each sample and their associated probability-to-exceed (PTE), respectively. In all cases we find that the model described in Section \ref{sec:theory} is able to describe the data with no evidence for a significant statistical tension. The top panel of Fig.\!~\ref{fig:bh} shows the normalised redshift distributions of the different bins used in this analysis, and can be used to visually assess the level of correlation between the different measurements.

      We see that, while we are able to measure $1-\bH$ to a reasonable accuracy ($\sim5-10\%$) in all of the \wisc~ redshift bins, the sensitivity for the 2MPZ sample is much poorer. This is not entirely unexpected: even though the cross-correlation between the Compton-$y$ map and the 2MPZ catalogue yields the highest signal-to-noise ratio of all the samples used here, the low-redshift range covered by this sample implies that the signal is strongly dominated by the 1-halo term. This can be seen in the top right panel of Fig.\!~\ref{fig:cls}. The constraining power of the 2MPZ cross-correlation degrades significantly since, as we have already described, we can only obtain reliable constraints on $1-\bH$ from the 2-halo contribution.
            
      Our results are in agreement with the estimate of a redshift-independent $1-\bH$ found by \cite{2019MNRAS.489..401Z} by calibrating cluster masses with CMB lensing ($1-\bH=0.71\pm0.10$), shown in Fig.\!~\ref{fig:bh} as a turquoise semi-transparent band. In turn, the grey band in the same figure shows the constraints on $1-\bH$ found by combining tSZ cluster counts and the $TT$ CMB power spectrum measured by \planck\ \citep{2016A&A...594A..24P} ($1-\bH=0.58\pm0.04$). This is the value of $1-\bH$ needed to simultaneously explain the amplitude of density perturbations predicted by the CMB and the abundance of massive clusters. This corresponds to a $\sim3-4\sigma$ tension with our results.

      We can however use our results to quantify whether approximating $1-\bH$ to be constant with redshift is supported by the data. To do so, we combine our six measurements under the assumption that they correspond to the same redshift-independent quantity. We do so by finding the quantity $\bar{b}_{\rm H}$ that minimises the $\chi^2$:
      \begin{equation}
        \chi^2=\sum_{i,j=1}^6 (b_{{\rm H},i}-\bar{b}_{\rm H}){\sf Cov}^{-1}_{b,ij}(b_{{\rm H},j}-\bar{b}_{\rm H}),
      \end{equation}
      where $b_{{\rm H},i}$ is the mass bias measured in the $i$-th redshift bin, and ${\sf Cov}_b$ is the covariance matrix of these measurements. Since the galaxy samples used in this analysis have significant redshift overlap, the off-diagonal elements of ${\sf Cov}_b$ cannot be ignored. We estimate ${\sf Cov}_b$ through jackknife resampling: we use the power spectra measured in each jackknife region described in Section \ref{ssec:methods.cov} to estimate the best-fit value of $b_{{\rm H},i}$ for each of them, and then calculate the covariance through Eq.\!~\ref{eq:cov_jk}. Since $\bar{b}_{\rm H}$ is a linear parameter, its best-fit and standard deviation can be found analytically as:
      \begin{equation}
        \bar{b}_{\rm H}=\frac{\sum_{ij}{\sf Cov}_{b,ij}^{-1}b_{{\rm H},i}}{\sum_{ij}{\sf Cov}^{-1}_{b,ij}},
        \hspace{12pt}
        \sigma(\bar{b}_{\rm H})=\left(\sum_{ij}{\sf Cov}^{-1}_{b,ij}\right)^{-1/2}.
      \end{equation}
      At this point it is worth acknowledging that this estimate of $\bar{b}_{\rm H}$ is only strictly consistent if the posterior distribution of the $b_{{\rm H},i}$ is Gaussian. We find that, with the exception of the 2MPZ sample, which has a comparatively small statistical power, the marginalised distributions in all redshift bins are sufficiently well-behaved that this is a reasonable approximation (e.g. see 1-dimensional distribution for $\bH$ in Fig.\!~\ref{fig:triangle}). Our combined constraint on $\bar{b}_{\rm H}$ following this procedure is:
      \begin{equation}
        1-\bar{b}_{\rm H}=0.75\pm0.03.
      \end{equation}
      More interestingly, the $\chi^2$ value associated with this measurement is $\chi^2=2.1$, which has an associated PTE $\simeq0.8$. We therefore find that the assumption of a constant mass bias with redshift is compatible with our measurements. This agrees with the results of \cite{2020Chiang}, who used the SZ effect to explore the cosmic thermal history. It also agrees with the results of \cite{2019A&A...626A..27S}, found using tSZ-selected clusters. The preferred value of $1-\bar{b}_{\rm H}$ found also agrees with the results of \cite{2018MNRAS.480.3928M} in cross-correlation with galaxies at very low redshift, and those of \cite{2019arXiv190707870M} using weak-lensing cross-correlations at higher $z$. 
      \begin{figure*}
        \centering
        \includegraphics[width=0.7\textwidth]{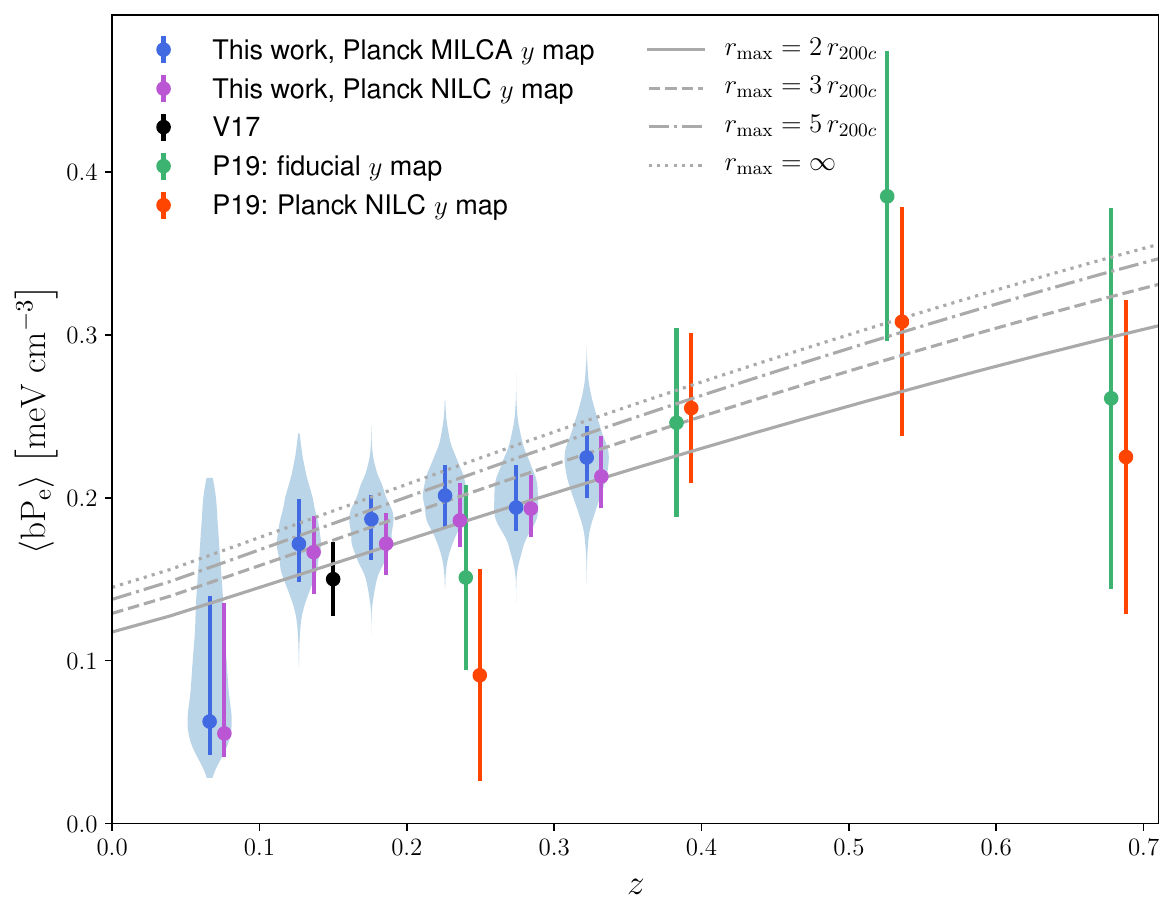}
        \caption{Summary figure showing our constraints on the bias-weighted thermal pressure $\langle bP_e\rangle$. Our measurements are shown as blue circles with error bars and a `violin' background. The circle and error bars show the peak of the 1D posterior distribution and 68\% confidence interval, while the violins show the full 1D posterior distribution. We also plot our measurements using the \planck\ NILC $y$ map, in violet. The grey lines show the predictions of the shock-heating model of \citet{2012ApJ...758...75B} for different values of the $r_{\rm max}$ threshold cluster radius (see legend). For comparison, the figure also shows the constraints on the same parameter found by \citet{2017MNRAS.467.2315V} (black) and \citet{2019arXiv190413347P} (red and green).}
        \label{fig:by}
      \end{figure*}
      
       While the best-fit value of $1-\bH\simeq0.8$ is in broad agreement with the expectation from hydrodynamical simulations \citep{2016ApJ...827..112B}, the estimate from CMB lensing mass calibration \citep{2019MNRAS.489..401Z}, and other direct calibration efforts \citep[e.g.][]{2016MNRAS.456L..74S,2019A&A...621A..40E}, which seem to prefer smaller missing mass fractions ($1-\bH\simeq0.8$), our results are in slight tension with the value of $1-\bH$ found by the \planck\ number counts analysis in combination with primary CMB anisotropies. A possible resolution of the mild tension between tSZ cluster counts and CMB data could be the existence of new physics modifying the late-time expansion and structure growth. However, most of these modifications (e.g. a one-parameter excursion where the dark energy equation of state takes a larger value $w\sim-0.7$) would make the existing tension in the value of the local expansion rate between CMB and local measurements \citep{2019ApJ...876...85R} worse. A careful study of all possibilities, invoking both non-standard physics and more sophisticated astrophysical models, together with improved datasets, is therefore necessary before this tension can be fully resolved.

    \subsubsection{Tomographic measurement of thermal gas pressure}\label{ssec:results.fid.bpe}
      So far we have presented our constraints in terms of a the mass bias parameter $1-\bH$, since this is the main source of systematic uncertainty in the cosmological analysis of cluster number counts carried out by \planck. However, the physical interpretation of this parameter (the fraction of missing mass estimated from X-ray measurements under the assumption of hydrostatic equilibrium) is not directly related to the physical process that allows us to constrain it through the cross-correlation of $y$ and $\delta_g$: the fact that galaxy density and pressure trace the same underlying dark-matter fluctuations. In this sense, a more direct observable is the bias-weighted pressure $\langle bP_e\rangle$. This quantity can be interpreted both as the relation between large-scale matter and pressure fluctuations and as the halo-bias-weighted thermal energy density of all haloes at a given redshift \citep{2017JCAP...11..040B}. It has been measured at low  redshifts by \cite{2017MNRAS.467.2315V} making use of galaxy groups, and at higher redshifts by the Dark Energy Science Collaboration \citep{2019arXiv190413347P}. The redshift range $0.1\lesssim z\lesssim0.4$, containing a large fraction of the tSZ sources detected by \planck\ \citep{2016A&A...594A..27P}, has been so far fairly unconstrained tomographically through this type of measurements.

      We derive constraints on $\langle bP_e\rangle$ from our data by reprocessing our Monte-Carlo chains, computing $\langle bP_e\rangle$ at each sample to find its 1-dimensional posterior distribution. The results are listed in Table \ref{tab:results} and shown in Fig.\!~\ref{fig:by}, together with our measurements using the \planck\ NILC $y$ map, and the measurements of \cite{2017MNRAS.467.2315V} and \cite{2019arXiv190413347P}. Our results are in good qualitative agreement with the trend of these previous measurements, as well as with the predictions of the shock heating models of \cite{2012ApJ...758...75B}. In these models, the thermal energy entering Eq.\!~\ref{eq:by} is estimated by integrating the pressure profile of \cite{2012ApJ...758...75B} up to a radius $r_{\rm max}=N\,r_{200c}$. As a visual aid to evaluate the agreement of our results with these models, the predictions for $N=2,\,3,\,5$ and $\infty$ are shown as solid, dashed, dot-dashed and dotted grey lines respectively in Fig.\!~\ref{fig:by}. These results are the most precise measurement of this quantity to date. The main factors that contribute to the improved constraining power are the larger amplitude of the tSZ signal towards low redshifts and the high density of tracers in the 2MPZ and \wisc\ samples.

  \subsection{Systematics analysis}\label{ssec:results.syst}
    \subsubsection{tSZ systematics}\label{sssec:results.syst.y}
      No component separation method is perfect, and the MILCA Compton-$y$ map used in our fiducial analysis is known to suffer from small levels of contamination from various other astrophysical components. The most relevant for this analysis is the presence of Galactic and extragalactic dust. As described in Section \ref{sssec:methods.syst.deproj}, we remove contamination from Galactic dust at the map level in both $y$ and $\delta_g$. The extragalactic component, the so-called Cosmic Infrared Background (CIB), is however a more relevant concern, given that it traces the large-scale structure, and is therefore statistically correlated with both of our observables ($\delta_g$ and $y$). Since the CIB is most relevant near the peak of star formation ($z\sim2$), we expect this contamination to be small, but it must be quantified carefully.
      
      To do so, we follow the same method used in \cite{2017MNRAS.467.2315V}. We model the CIB contamination in the $y$ map as:
      \begin{equation}
        y_{\rm obs}(\nv)=y_{\rm true}(\nv)+\epsilon_{\rm CIB}\,c(\nv),
      \end{equation}
      where $y_{\rm obs}$ and $y_{\rm true}$ are the observed and true Compton-$y$ maps, $\epsilon_{\rm CIB}$ is a free parameter, and $c(\nv)$ is a template for the CIB emission. For our analysis we used the 545 GHz map released by \planck\ as a proxy for CIB emission. The different cross-correlation between $y_{\rm obs}$ and both $\delta_g$ and $c$ are then given by:
      \begin{align}\label{eq:cl_yc}
        &C^{yc,{\rm obs}}_\ell= C^{yc}_\ell+\epsilon_{\rm CIB}\,C^{cc}_\ell,\\
        &C^{yg,{\rm obs}}_\ell= C^{yg}_\ell+\epsilon_{\rm CIB}\,C^{gc}_\ell,
      \end{align}
      where $C^{cc}_\ell$, $C^{yc}_\ell$,  and  $C^{gc}_\ell$ are the auto-correlation of the CIB, and its intrinsic cross-correlations with $y$ and $\delta_g$ respectively. Since $C^{gc}_\ell$ can be estimated directly by cross-correlating our galaxy overdensity maps with the 545 GHz map, the only remaining step to quantify the contamination to the $y$-$\delta_g$ correlation is to estimate $\epsilon_{\rm CIB}$. To do so, we measure the cross-correlation between the MILCA $y$ map and the 545 GHz map and fit a model of the form of Eq.\!~\ref{eq:cl_yc}, with a single free parameter $\epsilon_{\rm CIB}$, and $C^{yc}_\ell$ and $C^{cc}_\ell$ given by the best-fit models for the CIB auto- and cross-correlation provided by \cite{2014A&A...571A..30P,2016A&A...594A..23P}. As reported in \cite{2018PhRvD..97f3514A}, this procedure yields an estimate of the CIB contamination
      \begin{equation}
        \epsilon_{\rm CIB}=(2.3\pm6.6)\times10^{-7}\,({\rm MJy}/{\rm sr})^{-1}.
      \end{equation}
      For this exercise, in order to reduce the noise variance from the Galactic component of the 545 GHz map, the cross-correlations $C^{yc,{\rm obs}}_\ell$ and $C^{gc}_\ell$ were estimated using \planck's 20\% Galactic mask. Additionally, all uncertainties were estimated using the jackknife method described in Section \ref{ssec:methods.cov}.
      
      Using this measurement of $\epsilon_{\rm CIB}$ and the estimate of $C^{gc}_\ell$ from the cross-correlation with the 545 GHz map, Fig.\!~\ref{fig:cls} shows, in the right panels, the estimated level of contamination from the CIB in our $y$-$\delta_g$ cross-correlation. In all cases the contamination is small, at the level of $\sim1\%$ of the signal, and therefore can be neglected in our analysis.

      As an additional test for systematics, we have also repeated our full analysis replacing the MILCA $y$ map with the NILC map released by \planck. Different component-separation methods are sensitive to different types of contamination, and this exercise is therefore both a necessary consistency check and a way to reinforce our conclusion that the level of CIB contamination is negligible. The resulting measurements of $1-\bH$ are shown in Fig.\!~\ref{fig:bh} next to the fiducial measurements, as grey circles with error bars. The results are in very good agreement with our fiducial analysis.
      
    \subsubsection{Photometric redshift uncertainties}\label{sssec:results.syst.pz}
      We have quantified the impact on our results of the uncertainties on the redshift distributions of the different samples used here by introducing the free width parameter $w_z$ with a 20\% top-hat prior. In order to study the impact of these uncertainties on our results, we have recomputed the constraints on $1-\bH$ fixing $w_z$ to its fiducial value of 1. The results are shown in Fig. \ref{fig:bh} as red downward-pointing triangles. We observe that the only effect of allowing $w_z$ to vary is to increase the final uncertainties on $1-\bH$ by about $20\%$. The final best-fit value of $1-\bH$ does not change significantly, and the posterior distribution of $w_z$ is always peaked around the fiducial value of $1$ (see e.g. bottom right panel of Fig. \ref{fig:triangle}).
      
    \subsubsection{Mass function parameterisation}\label{sssec:results.syst.mf}
      \begin{figure}
        \centering
        \includegraphics[width=0.49\textwidth]{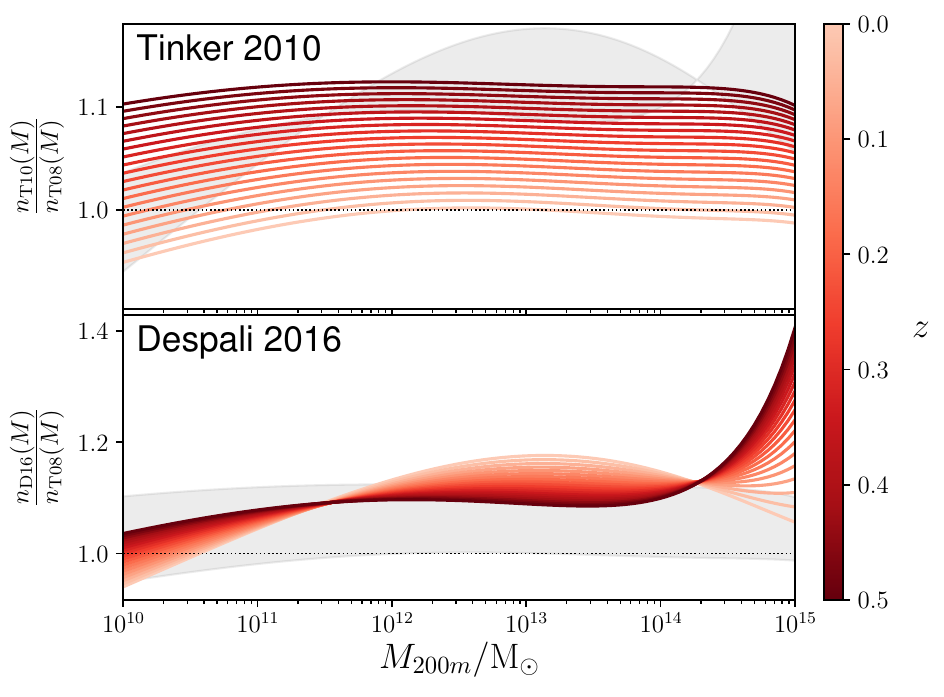}
        \caption{Ratio between the mass function parameterisations of \citet{2010ApJ...724..878T} and \citet{2016MNRAS.456.2486D} with the parameterisation of \citet{2008ApJ...688..709T} in the range of redshifts relevant to this analysis. A systematic offset of about $10$-$20\%$ between them can be observed, which causes the lower value of $1-b_{\rm H}$ found using the \citet{2010ApJ...724..878T} parameterisation (see Fig.\!~\ref{fig:bh}).}
        \label{fig:mf_ratio}
      \end{figure}
      Another source of systematics in our measurement is the theoretical uncertainty due to our choice of halo mass function in the halo model. Our fiducial mass function is the parameterisation of \cite{2008ApJ...688..709T}, which we use in order to be able to make a direct comparison with the results of \cite{2016A&A...594A..24P}. To explore the dependence on the choice of mass function parameterisation, we perform our analysis again using the updated parameterisation of \cite{2010ApJ...724..878T}. Fig.\!~\ref{fig:bh} shows the final constraints on $1-\bH$ in this case as burgundy squares, which are consistently lower by $\sim 2\sigma$. We find that this is caused by a redshift-dependent systematic offset between the two mass function parameterisations which reaches the level of $10$-$20\%$ in the range $0<z<0.4$. This is shown explicitly in Fig. \ref{fig:mf_ratio}, where it is evident that this offset is present for two mass function parameterisations: \citet{2010ApJ...724..878T} and \citet{2016MNRAS.456.2486D}. Since the relative differences of both models with that of \citet{2008ApJ...688..709T} are of the same order ($\sim10\%$ enhancement), we expect the constraints on $1-\bH$ using \citet{2016MNRAS.456.2486D} to be similar to those shown in Figure \ref{fig:bh} as burgundy squares. We have not explored any offsets potentially caused by our choice of halo bias parameterisation (having adopted the one described in \citet{2010ApJ...724..878T} throughout). Since the aim of this paper is not to characterise the theoretical uncertainties on the combination of halo mass function and halo bias, we leave the study of this issue for future work, and limit ourselves to using the parameterisation of \cite{2008ApJ...688..709T} in order to match previous analyses \citep[e.g.][]{2016A&A...594A..24P,2018MNRAS.477.4957B,2018MNRAS.473.4318A,2019MNRAS.489..401Z,2019arXiv190707870M}. It should be pointed out, however, that these systematics should not affect our results of the halo model bias for the Compton-$y$ parameter, $\langle bP_e\rangle$, as the latter is mainly affected by the relative amplitudes of the $\delta_g\times \delta_g$ and $y\times \delta_g$ power spectra, and is therefore robust to the mass function we use.

    \subsubsection{Galaxy clustering model}\label{sssec:results.syst.gc}
      To verify that our results are robust against the details of the HOD model used to parameterise the galaxy-matter connection, we have repeated the analysis for extended versions of our baseline model. We explore two such extensions:
      \begin{itemize}
        \item We include the threshold width for centrals $\sigma_{\rm lnM}$ as an additional free parameter with a broad top-hat prior, instead of fixing it to the value used by \cite{2018MNRAS.473.4318A}.
        \item We decouple central and satellite galaxies, allowing for haloes without a central galaxy to contain satellites. This implies removing the factor $N_c(M)$ in Eq. \ref{eq:hod1} and introducing $M_0$ (the mass threshold to have satellites) as a new free parameter that is not linked to $M_{\rm min}$.
      \end{itemize}
      In both cases we have confirmed that the constraints derived on $1-\bH$ do not deviate significantly from our fiducial results, and that our data are not able to constrain the new free HOD parameters introduced by each extension ($\sigma_{\rm lnM}$ and $M_0$ respectively). We therefore conclude that our results are insensitive to the specifics of the model used to characterise the clustering of galaxies, which is well described in the range of scales studied here by our fiducial 2-parameter HOD model .

\section{Conclusion}\label{sec:conclusion}
  Cross-correlating maps of the tSZ Compton-$y$ parameter with tomographic measurements of the projected galaxy distribution allows us to study the redshift evolution of the thermal gas pressure, since we expect both pressure and galaxies to trace the same underlying matter inhomogeneities. We have measured the $\delta_g$-$y$ cross-correlation to very high significance using public $y$ maps made available by the \planck\ collaboration \citep{2016A&A...594A..22P} and the 2MPZ and \wisc~ galaxy catalogues \citep{2014ApJS..210....9B,2016ApJS..225....5B} in six photometric redshift bins covering the redshift range $z\lesssim0.4$.
  
  Combining this measurement with a measurement of the galaxy auto-correlation allows us to break the degeneracy between the two parameters that relate $\delta_g$ and $y$ to the matter fluctuations: the galaxy bias, which we model effectively using a 2-parameter HOD prescription, and the mass bias parameter $1-\bH$; and thus enables us to make a redshift-dependent measurement of $1-\bH$. The results, shown in Table \ref{tab:results} and Fig.\!~\ref{fig:bh}, agree well with the calibration of tSZ cluster masses with CMB lensing measurements \citep{2019MNRAS.489..401Z}, and with similar measurements of $1-\bH$ made by \cite{2018MNRAS.480.3928M} through the same type of cross-correlation at lower redshifts. Nevertheless, we observe a slight tension with the measurements of $1-\bH$ made by \planck\ through the combination of cluster number counts and CMB primary anisotropies \citep{2016A&A...594A..24P}. More importantly, our tomographic measurement allows us to study the possible redshift dependence of the mass bias, an important ingredient in the cosmological analysis of cluster abundances. Within our uncertainties, we do not find any statistical evidence for a redshift dependence of $1-\bH$, in agreement with previous analyses \citep{2019A&A...626A..27S}.
  
  Perhaps more interestingly, our measurements can be interpreted as constraints on the bias-weighted mean gas pressure $\langle bP_e\rangle$, the equivalent of the large-scale galaxy bias for the $y$ parameter. This quantity is directly related to the energetics of gas in haloes, and can be used to constrain different heating models. Our results, shown in Table \ref{tab:results} and Fig.\!~\ref{fig:by}, agree well with previous measurements of the same quantity \citep{2017MNRAS.467.2315V,2019arXiv190413347P}, as well as with shock-heating models \citep{2012ApJ...758...75B}. This result is also the most precise measurement of the large-scale correlation between $\delta_g$ and $y$ to date.

  The main sources of systematic uncertainty in our analysis are contamination of the galaxy clustering auto-correlation by stars, Galactic dust and other observing conditions, the contamination from CIB emission in the $y$ map, and uncertainties in the galaxy redshift distribution due to the use of photometric redshifts. We address the galaxy clustering systematics by deprojecting templates for dust and star contamination at the map level, by deprojecting the expected contamination from zero-point fluctuations in the SuperCOSMOS photographic plates at the power spectrum level, and by masking out the largest scales ($\ell<10$) in the galaxy clustering auto-correlation. We use the 545 GHz \planck\ map as a tracer of CIB to quantify the level of contamination in the $y$-$\delta_g$ cross-correlation, and find it to be negligible (as could be expected given the relatively low redshift of our sample compared with the peak of star formation, $z\sim2$). Finally, we determine that the most relevant form of systematic associated with uncertainties in the redshift distributions is that associated with the width of $p(z)$. To address this, we add a new parameter to our model, $w_z$, which describes the distribution widths. We marginalise over $w_z$ with a 20\% prior, which we have determined to be significantly larger than the expected uncertainty on the true width. This has the effect of degrading our constraints on $1-\bH$ and $\langle bP_e\rangle$ by $\sim20\%$ without modifying the best-fit value of either quantity significantly.
  
  We must also note that our parameter constraints have been derived for a fixed cosmological model, corresponding to the best-fit parameters found by \planck\ \citep{2018arXiv180706209P}. The most significant consequence is the fact that our constraints are strongly degenerate with the amplitude of matter fluctuations, parameterised by $\sigma_8$. Therefore, the agreement between our constraints on $1-\bH$ and those found by \planck\ \citep{2016A&A...594A..24P} can be thought of as a proof of the consistency between the properties of the galaxy distribution in 2MPZ and \wisc~ and the CMB anisotropies. However, our results regarding the redshift evolution of $1-\bH$ (or lack thereof), and the agreement of $\langle bP_e\rangle$ with existing heating models, are only possible thanks to the tomographic cross-correlation with the galaxy density fluctuations. Robust joint constraints on cosmological and astrophysical parameters could be achieved by a combined analysis of galaxy clustering, Compton-$y$ maps and gravitational lensing data (either from cosmic shear or CMB lensing observations). We leave this analysis for future work.
  
  The scientific yield of these types of observations will increase significantly with data from current and near-future experiments, such the Advanced Atacama Cosmology Telescope \citep{2016SPIE.9910E..14D} or the Simons Observatory \citep{2019JCAP...02..056A} on the CMB side, and the Large Synoptic Survey Telescope \citep{2009arXiv0912.0201L} or the Euclid satellite \citep{2011arXiv1110.3193L} in terms of galaxy clustering and cosmic shear. The main advances will come in the form of lower-noise and higher-resolution $y$ and CMB lensing maps, as well as a dense ($\sim30\,{\rm arcmin}^{-2}$) sampling of the galaxy distribution to much higher redshifts ($z\lesssim2$). This increase in sensitivity, however, will have to be accompanied by a much better control of systematics such as contamination from CIB and other sources, galaxy clustering systematics or photo-$z$ uncertainties.

\section*{Data Availability}
  \begin{itemize}
      \item The Compton-$y$ maps are publicly available by the \planck\ collaboration \citep{2016A&A...594A..22P}.
      \item The low-redshift photometric redshift catalogues (2MASS Photometric Redshift catalogue, WISE $\times$ SuperCOSMOS catalogue) are outlined in \cite{2014ApJS..210....9B} and \cite{2016ApJS..225....5B}, respectively.
  \end{itemize}

\section*{Acknowledgements}
  We would like to thank Boris Bolliet and Eiichiro Komatsu for useful comments and discussions. NK is funded by the Science and Technology Facilities Council (STFC). DA acknowledges support from the Beecroft trust and from STFC through an Ernest Rutherford Fellowship, grant reference ST/P004474/1. MB is supported by the Polish Ministry of Science and Higher Education through grant DIR/WK/2018/12, and by the Polish National Science Center through grant no. 2018/30/E/ST9/00698. JAP was supported by the European Research Council under grant no. 670193.
  
  The current manuscript is a revised version of the one originally posted on {\tt arXiv} and published in \textit{Monthly Notices of the Royal Astronomical Society} (\cite{2019MNRAS...491..5456}) after a bug in our analysis pipeline was brought to our attention by Ruy Makiya, whom we would like to sincerely thank.
  
  The {\tt python} packages {\tt healpy}, {\tt numpy} and {\tt scipy} were used for data analysis, and {\tt matplotlib} was used to plot our results.
  
\setlength{\bibhang}{2.0em}
\setlength\labelwidth{0.0em}
\bibliography{paper}

\end{document}